\documentclass[apj,iop]{emulateapj}
\usepackage{acronym,epstopdf,graphicx,hyperref,textcomp}

\begin{document}

\title{Searches for continuous gravitational waves from nine young supernova
remnants}

\author{%
J.~Aasi$^{1}$,
B.~P.~Abbott$^{1}$,
R.~Abbott$^{1}$,
T.~Abbott$^{2}$,
M.~R.~Abernathy$^{1}$,
F.~Acernese$^{3,4}$,
K.~Ackley$^{5}$,
C.~Adams$^{6}$,
T.~Adams$^{7,8}$,
P.~Addesso$^{9}$
% ...
}

\address {$^{1}$LIGO, California Institute of Technology, Pasadena, CA 91125, USA }
\address {$^{2}$Louisiana State University, Baton Rouge, LA 70803, USA }
\address {$^{3}$Universit\`a di Salerno, Fisciano, I-84084 Salerno, Italy }
\address {$^{4}$INFN, Sezione di Napoli, Complesso Universitario di Monte Sant'Angelo, I-80126 Napoli, Italy }
\address {$^{5}$University of Florida, Gainesville, FL 32611, USA }
\address {$^{6}$LIGO Livingston Observatory, Livingston, LA 70754, USA }
\address {$^{7}$Cardiff University, Cardiff, CF24 3AA, United Kingdom }
\address {$^{8}$Laboratoire d'Annecy-le-Vieux de Physique des Particules (LAPP), Universit\'e de Savoie, CNRS/IN2P3, F-74941 Annecy-le-Vieux, France }
\address {$^{9}$University of Sannio at Benevento, I-82100 Benevento, Italy and INFN, Sezione di Napoli, I-80100 Napoli, Italy }
% ...

\begin{abstract}
We describe directed searches for continuous gravitational waves in data from
the sixth LIGO science data run.
The targets were nine young supernova remnants not associated with pulsars;
eight of the remnants are associated with non-pulsing suspected neutron stars.
One target's parameters are uncertain enough to warrant two searches, for a
total of ten.
Each search covered a broad band of frequencies and first and second frequency
derivatives for a fixed sky direction.
The searches coherently integrated data from the two LIGO interferometers over
time spans from 5.3--25.3~days using the matched-filtering
$\mathcal{F}$-statistic.
We found no evidence of gravitational-wave signals.
We set 95\% confidence upper limits as strong (low) as $4\times10^{-25}$ on
intrinsic strain, $2\times10^{-7}$ on fiducial ellipticity, and
$3\times10^{-6}$ on $r$-mode amplitude.
These beat the indirect limits from energy conservation and are within the
range of theoretical predictions for neutron-star ellipticities and $r$-mode
amplitudes.
\end{abstract}


\maketitle

\acrodef{CCO}{central compact object}
\acrodef{GW}{gravitational wave}
\acrodef{LHO}{LIGO Hanford Observatory}
\acrodef{LLO}{LIGO Livingston Observatory}
\acrodef{LIGO}{the Laser Interferometer Gravitational-wave Observatory}
\acrodef{LSC}{LIGO Scientific Collaboration}
\acrodef{PSD}{power spectral density}
\acrodef{PWN}{pulsar wind nebula}
\acrodefplural{PWN}[PWNe]{pulsar wind nebulae}
\acrodef{S1}{its first science run}
\acrodef{S5}{the fifth LIGO science run}
\acrodef{S6}{the sixth LIGO science run}
\acrodef{SFT}{Short Fourier Transform}
\acrodef{SNR}{supernova remnant}
\def\fscan{Fscan}

\section{Introduction}

Young neutron stars are attractive targets for searches for continuous
\acp{GW} even if they are not detected as pulsars \citep{Wette2008, Owen2009,
S5CasA}.
Some are seen as non-pulsing \acp{CCO} in \acp{SNR}, and some young \acp{PWN}
and \acp{SNR} indicate the location of a young neutron star with enough
precision for a directed search---a search over frequency and spin-down
parameters, but not over sky positions.
Some young pulsars spin fast enough to emit \acp{GW} in the frequency band of
ground-based interferometers such as \ac{LIGO} and Virgo, and therefore some
young non-pulsars may also spin fast enough.
Even without observed pulsations and spin-down parameters, it is possible to
estimate an indirect upper limit on \ac{GW} emission, analogous to the
spin-down limit \citep{Shklovskii1969} for known pulsars, based on the age of
and distance to the star plus energy conservation \citep{Wette2008}.
Given the great uncertainties in predictions of \ac{GW} emission from young
neutron stars, we use this indirect limit rather than those predictions to
pick targets for directed searches.

We describe such searches of data from \ac{S6} for continuous \acp{GW} from
Cas~A and eight more supernova remnants with known or suspected young isolated
neutron stars with no observed electromagnetic pulsations.
These targets were chosen so that a computationally feasible coherent search
similar to \citet{S5CasA} could beat the indirect limits on \ac{GW} emission.
Therefore each search had a chance of detecting something, and non-detections
could constrain the star's \ac{GW} emission, provided that emission is at a
frequency in the band searched.
No search found evidence for a \ac{GW} signal, and hence the main result is a
set of upper limits similar to those presented in \citet{S5CasA}.
These upper limits on \ac{GW} emission translate into upper limits on the
fiducial ellipticity and $r$-mode amplitude of each neutron star as a function
of \ac{GW} frequency the star could be emitting (see
Subsection~\ref{ss:results}).
The ellipticity and $r$-mode upper limits set by the searches described here
were within the ranges of theoretical predictions
\citep{Johnson-McDaniel2013, Bondarescu2009}, another indicator that these
searches reached interesting sensitivites (see Section~\ref{s:disc}).

For context, we compare to the other continuous \ac{GW} searches, which
correspond to three other astronomical populations that nonetheless share
astrophysical emission mechanisms and other properties \citep{Owen2009}.
Directed searches occupy a middle ground between all-sky searches
and targeted searches for known pulsars in the key tradeoff for continuous
waves:
Searches with greater sensitivity and less computational cost require more
astronomical information, and have different indirect limits to beat to reach
an interesting sensitivity.

The first search for continuous waves in LIGO data, from \ac{S1}, was for a
single known pulsar \citep{S1pulsar}.
Such a search, guided by a precise timing solution, is computationally cheap
and achieves the best strain sensitivity for a given amount of data since all
available data can be integrated coherently.
Since then, searches of data up to \ac{S6} have targeted up to 195 pulsars
\citep{S2pulsars, S3S4pulsars, S5Crab, VirgoVela, S5pulsars, S6pulsars}.
The four most recent of these papers set direct upper limits on \ac{GW}
emission stricter than the spin-down limits derived from energy conservation,
for a few of the pulsars searched, thereby marking the point at which
\ac{LIGO} and Virgo began revealing new information about these pulsars.
The upper limits also corresponded to neutron-star ellipticities within the
range of theoretical predictions for exotic equations of state
\citep{Owen2005}.

Other continuous \ac{GW} searches have surveyed the whole sky for neutron
stars not seen as pulsars, using great computational power to cover wide
frequency bands and large ranges of spin-down parameters \citep{S2Hough,
S2Fstat, S4PSH, S4Einstein, S5PowerFlux, S5Einstein, S5PowerFlux2,
S5Einstein2, S5Hough, VirgoAllSky} and recently possible binary parameters too
\citep{AllSkyBinary}.
Several of the recent all-sky searches have set direct upper limits
competitive with indirect upper limits based on simulations of the galactic
neutron-star population \citep{KnispelAllen}.

Between these two extremes of computational cost and sensitivity are the
directed searches, where the sky location (and thus the detector-frame Doppler
modulation) is known but the frequency and other parameters are not.
Directed searches can be divided further into searches for isolated
neutron stars (the type of search described in this paper), and searches for
neutron stars in binary systems, with particular emphasis on accreting neutron
stars in close (low mass x-ray) binaries.
For accreting neutron stars, a different indirect limit can be set based on
angular momentum conservation \citep{Papaloizou1978}.
Unlike the energy conservation-based indirect limits for other neutron star
populations, there is an argument (partially based on observations) that
accreting neutron stars emit close to their limit, which also corresponds to
reasonable ellipticities and $r$-mode amplitudes \citep{Bildsten1998}.
So far the only accreting neutron star targeted has been the one in the
low-mass X-ray binary Sco~X-1 \citep{S2Fstat, S4radiometer,
S5Stochastic, S5Sideband}.
Searches for this object must cover not only a range of \ac{GW} frequencies
since no pulsations are observed, but also a range of orbital parameters since
there are substantial uncertainties in these.
Direct upper limits from searches for Sco~X-1 have not beaten the indirect
limit derived from accretion torque balance \citep{Papaloizou1978}, but may
with data from interferometers upgraded to the ``advanced'' sensitivity
\citep{aLIGO, AdvancedLIGO, Sammut2014}.

The type of directed search described here, for isolated neutron stars not
seen as pulsars, was first performed on data from \ac{S5} for the \ac{CCO} in
the \ac{SNR} Cas~A \citep{S5CasA}.
Since then similar searches, using different data analysis methods, have been
performed for supernova 1987A and unseen stars near the galactic center
\citep{S5Stochastic, S5GalacticCenter}.
Directed searches for isolated neutron stars are intermediate in cost and
sensitivity between targeted pulsar searches and all-sky searches because a
known sky direction allows for searching a wide band of frequencies and
frequency derivatives with much less computing power than the all-sky
wide-band searches \citep{Wette2008} and no search over binary parameters is
needed.
The indirect limits to beat are numerically similar to those for known
pulsars---the strain limit for Cas~A is almost identical to that for the Crab
pulsar.
One disadvantage this type of search has compared to pulsar searches is that
the spin frequencies of the neutron stars are not known.
Based on pulsar statistics, it is likely that most of these stars are not
spinning fast enough to be emitting \acp{GW} in the detectable frequency band,
making it all the more important to search multiple targets.
Here we improve the methods of the S5 Cas~A search \citep{S5CasA} and extend
our search targets to nine young supernova remnants total.

The rest of this article is structured as follows:
In Sec.~\ref{s:searches} we present the methods, implementation, and
results of the searches.
The upper limits set in the absence of evidence for a signal are presented in
Sec.~\ref{s:uls}, and the results are discussed in Sec.~\ref{s:disc}.
In the Appendix we describe the performance of the analysis pipeline on
hardware injected signals.

\section{Searches}
\label{s:searches}

\begin{table}
\begin{center}
\caption{
\label{t:targets}
Target objects and astronomical parameters used in each search
}
\begin{tabular}{rllcc}
\hline
\hline
SNR & Other name & RA+dec & $D$ & $a$ \\
(G name) & & (J2000) & (kpc) & (kyr) \\
\tableline
1.9$+$0.3 & & 174846.9$-$271016 & 8.5\hspace{0.5em} & 0.1\hspace{0.5em}
\\
18.9$-$1.1 & & 182913.1$-$125113 & 2\hspace{1.25em} & 4.4\hspace{0.5em}
\\
93.3$+$6.9 & DA 530 & 205214.0$+$551722 & 1.7\hspace{0.5em} & 5\hspace{1.25em}
\\
111.7$-$2.1 & Cas A & 232327.9$+$584842 & 3.3\hspace{0.5em} &
0.3\hspace{0.5em}
\\
189.1$+$3.0 & IC 443 & 061705.3$+$222127 & 1.5\hspace{0.5em} & 3\hspace{1.25em}
\\
266.2$-$1.2 & Vela Jr.\ & 085201.4$-$461753 & 0.2\hspace{0.5em} & 0.69
\\
266.2$-$1.2 & Vela Jr.\ & 085201.4$-$461753 & 0.75 & 4.3\hspace{0.5em}
\\
291.0$-$0.1 & MSH 11$-$62 & 111148.6$-$603926 & 3.5\hspace{0.5em} &
1.2\hspace{0.5em}
\\
347.3$-$0.5 & & 171328.3$-$394953 & 0.9\hspace{0.5em} & 1.6\hspace{0.5em}
\\
350.1$-$0.3 & & 172054.5$-$372652 & 4.5\hspace{0.5em} & 0.6\hspace{0.5em}
\\
\hline
\end{tabular}
\end{center}
Values of distance $D$ and age $a$ are at the optimistic (nearby and young)
end of ranges given in the literature, except for the second search for
Vela~Jr.
See text for details and references.
\end{table}

\subsection{Data selection}

\begin{table*}
\begin{center}
\caption{
\label{t:targets2}
Derived parameters used in each search
}
\begin{tabular}{rrrrrrrrr}
\hline
\hline
SNR & $f_{\min}$ & $f_{\max}$ & $T_\mathrm{span}$ & $T_\mathrm{span}$ &
Start of span & H1 & L1 & Duty
\\
(G name) & (Hz) & (Hz) & (s) &
(days) & (UTC, 2010) & SFTs & SFTs & factor
\\
\tableline
1.9$+$0.3 & 141 & 287 & 788\,345 & 9.1 & Aug 22 00:23:45 & 356 & 318 & 0.77
\\
18.9$-$1.1 & 132 & 298 & 2\,186\,572 & 25.3 & Aug 13 02:02:24 & 786 & 912 &
0.70 
\\
93.3$+$6.9 & 109 & 373 & 2\,012\,336 & 23.3 & Aug 10 18:49:49 & 770 & 813 &
0.71
\\
111.7$-$2.1 & 91 & 573 & 730\,174 & 8.4 & Aug 22 10:27:49 & 332 & 289 & 0.77
\\
189.1$+$3.0 & 101 & 464 & 1\,553\,811 & 18.0 & Aug 13 07:55:32 & 650 & 634 &
0.74
\\
266.2$-$1.2 & 46 & 2034 & 456\,122 & 5.3 & Jul 30 06:17:12 & 218 & 186 & 0.80
\\
266.2$-$1.2 & 82 & 846 & 1\,220\,616 & 14.1 & Aug 17 02:58:47 & 525 & 503 &
0.76
\\
291.0$-$0.1 & 124 & 315 & 1\,487\,328 & 17.2 & Aug 14 00:53:35 & 629 & 615 &
0.75
\\
347.3$-$0.5 & 82 & 923 & 903\,738 & 10.5 & Aug 20 22:00:05 & 397 & 370 & 0.76
\\
350.1$-$0.3 & 132 & 301 & 1\,270\,309 & 14.7 & Aug 16 13:10:34 & 538 & 519 &
0.75
\\
\hline
\end{tabular}
\end{center}
The span reported is the final one, including the possible extension to the
end of an \protect\ac{SFT} in progress at the end of the originally requested
span.
The duty factor reported is total \protect\ac{SFT} time divided by
$T_\mathrm{span}$ divided by the number of interferometers (two).
\end{table*}

\ac{S6} ran from July 7 2009 21:00:00 UTC (GPS 931035615) to October 21 2010
00:00:00 UTC (GPS 971654415).
It included two interferometers with 4-km arm lengths, H1 at \ac{LHO} near
Hanford, Washington and L1 at \ac{LLO} near Livingston, Louisiana.
It did not include the 2-km H2 interferometer that was present at \ac{LHO}
during earlier runs.
Plots of the noise \ac{PSD} curves and descriptions of the improvements over
S5 can be found, for example, in \citet{Aasi:2014usa}.
A description of the calibration and uncertainties can be found in
\citet{Bartos2011}.
The phase calibration errors at the frequencies searched were up to 7$^\circ$
and 10$^\circ$ for H1 and L1 respectively, small enough not to affect the
analysis.
The corresponding amplitude calibration errors were 16\% and 19\%
respectively.
For reasons discussed in \citet{S6pulsars} we estimate the maximum amplitude
uncertainty of our joint H1-L1 results to be 20\%.

Concurrently with the \ac{LIGO} \ac{S6} run, the Virgo interferometer near
Cascina, Italy had its data runs VSR2 and VSR3.
Although Virgo noise performance was better than \ac{LIGO} in a narrow band
below roughly 40~Hz, it was worse than \ac{LIGO} by a factor 2--3 in amplitude
at the higher frequencies of the searches described here.
Virgo's declination response function for many-day observations
\relax{averaged over inclinations and polarizations} is within about 10
percent of that of \ac{LHO}, \relax{and even extreme inclinations and
polarizations are not too far from average} [see Fig.~4 \relax{and Eq.~(86)
respectively} of \citet{Jaranowski1998}].
\relax{Hence Virgo's} single-interferometer sensitivity is worse by a factor
2--3 \relax{in amplitude, and since the signal-to-noise is added in
quadrature between interferometers, the addition of Virgo would enhance the
sensitivity to a typical source by at most a few percent---much less than the
LIGO calibration uncertainty}.
Since data analysis costs the same for all interferometers and computational
resources are limited, the searches described here used only \ac{LIGO} data.

Like many other continuous-wave searches, those reported here used \ac{GW}
data in the \ac{SFT} format.
The series of science-mode data, interrupted by planned (maintenance) and
unplanned downtime (earthquakes etc.), minus short segments which were
``category 1'' vetoed \citep{Aasi:2014usa}, was broken into segments of
$T_\mathrm{SFT} = 1800$~s.
There were a total of 19\,268 of these segments for H1 and L1 during the
\ac{S6} run.
Each 30-minute segment was first high pass filtered in the time domain through
a tenth order Butterworth filter with a knee frequency of 30~Hz to attenuate
low frequency seismic noise.
Then it was Tukey windowed with parameter 0.001 (i.e.\ only 0.1\% of samples
were modified) to mitigate edge artifacts.
Finally each segment was Fourier transformed and frequencies from 40--2035~Hz
were recorded in the corresponding \ac{SFT}.

Although a directed search is computationally more tractable than an all-sky
search, computational costs nonetheless restricted us to searching a limited
time span $T_\mathrm{span}$ of the \ac{S6} data.
This span, and the frequency band $f_{\min}$--$f_{\max}$, were determined for
each target by an algorithm designed to fix the computational cost per target
as described in Subsection~\ref{ss:parameters}.
The data selection criterion was the same as in \citet{S5CasA}, maximizing the
figure of merit
\begin{equation}
\label{data-fom}
\sum_{f,t} \frac{1}{S_h(f,t)}
\end{equation}
where the sums run over the given $T_\mathrm{span}$, $f_{\min}$, and
$f_{\max}$ for each target.
Here $f$ is the frequency of each bin (discretized at $1/T_\mathrm{SFT}$), $t$
is the time stamp of each \ac{SFT}, and $S_h$ is the strain noise \ac{PSD}
harmonically averaged over the H1 and L1 interferometers.
Maximizing this figure of merit roughly corresponds to optimizing (minimizing)
the detectable \ac{GW} strain, harmonically averaged over the frequency band.
Although the frequency band for each search varied target by target, the
sum was dominated by the least noisy frequencies that are searched for all
targets, and thus the optimization always picked time spans near the end of
\ac{S6} when the noise at those frequencies was best (least) and the \ac{SFT}
duty factor (total \ac{SFT} time divided by $T_\mathrm{span}$ divided by
numbers of interferometers [two]) was highest.
This figure of merit also neglects the small effect where \ac{LHO} is better
for high declination sources and \ac{LLO} is better for low
\citep{Jaranowski1998}.
Since the optimal data stretches tended to have comparable amounts of H1 and
L1 data, the declination effect was at most a few percent, less than the
amplitude calibration uncertainties.

\subsection{Analysis method}

The analysis was based on matched filtering, the optimal method for detecting
signals of known functional form.
To obtain that form we assumed that the instantaneous frequency of the
continuous (sinusoidal) \acp{GW} in the solar system barycenter was
\begin{equation}
\label{ft}
f(t) \simeq f + \dot{f}(t-t_0) + \frac{1}{2} \ddot{f} (t-t_0)^2.
\end{equation}
That is, we assumed that none of the target neutron stars glitched (had abrupt
frequency jumps) or had significant timing noise (additional, perhaps
stochastic, time dependence of the frequency) during the observation.
We also neglected third and higher derivatives of the \ac{GW} frequency, based
on the time spans and ranges of $\dot{f}$ and $\ddot{f}$ covered.
The precise expression for the interferometer strain response $h(t)$ to an
incoming continuous \ac{GW} also includes amplitude and phase modulation by
the changing of the beam patterns as the interferometer rotates with the
earth.
It depends on the source's sky location and orientation angles, as well as on
the parameters of the interferometer, and takes the form of four sinusoids.
We do not reproduce the lengthy expression here, but it can be found in
\citet{Jaranowski1998}.

The primary detection statistic was the multi-interferometer
$\mathcal{F}$-statistic \citep{Cutler2005}.
This is based on the single-interferometer $\mathcal{F}$-statistic
\citep{Jaranowski1998}, which combines the results of matched filters for the
four sinusoids of the signal in a way that is computationally fast and nearly
optimal \citep{Prix2009}.
In Gaussian noise $2\mathcal{F}$ is drawn from a $\chi^2$ distribution with
four degrees of freedom, and hence $\mathcal{F}/2$ is roughly a power
signal-to-noise ratio.

We used the implementation of the $\mathcal{F}$-statistic in the LALSuite
package, tag \texttt{S6SNRSearch}, publicly available at
\texttt{https://www.lsc-group.phys.uwm.edu/\linebreak[0]daswg/\linebreak[0]projects/\linebreak[0]lalsuite.html}.
In particular most of the computing power of the search was spent in the
\texttt{ComputeFStatistic\_v2\_SSE} program, which unlike the version used in
the preceding search of this type \citep{S5CasA} uses the Intel SSE2
floating-point extensions and only 8 terms rather than 16 in the Dirichlet
kernel.
Both of these changes sped up the analysis (see below).

The algorithm for setting up a ``template bank,'' or choosing discrete points
in the parameter space of $(f,\dot{f},\ddot{f})$ to search, was the same as in
\citet{S5CasA}.
The ``mismatch'' or maximum loss of $2\mathcal{F}$ due to discretization of
the frequency and derivatives \citep{Owen1996, Brady1998} was 0.2, again the
same as in \citet{S5CasA}.
Choosing to keep the computational cost the same for all searches resulted in
some variation of the total number of templates per search,
3--12$\times10^{12}$ compared to the $7\times10^{12}$ in \citet{S5CasA}.

\subsection{Target objects}

The goal of these searches was to target young non-pulsing neutron stars.
Starting with the comprehensive catalog of \acp{SNR} \citep{Green2009,
Green2014}, augmented by a search of the recent literature, we narrowed
the list to remnants with confirmed associated non-pulsing ``point''
sources---\acp{CCO} (mostly soft thermal emission, no radio, sub-arcsecond
size) or small \acp{PWN} (mostly hard nonthermal emission, sub-arcminute
size) or candidates.
These strongly indicate the presence of a neutron star, although they do not
indicate if it is emitting in the \ac{LIGO} frequency band.
We make our searches in the hopes that some of the stars are emitting in this
band.
We also included \ac{SNR}~G1.9+0.3, although a point source is not visible
(and may not exist since the supernova may have been Type~Ia), because this
remnant is the youngest known and is small enough to search with a single sky
location.

The final selection of target objects and search parameters was based on
beating the indirect upper limit on \ac{GW} emission due to energy
conservation.
This upper limit is based on the optimistic assumption that all of the star's
(unobserved) spin-down is due to \ac{GW} emission, and has been since the
supernova.
In terms of the ``intrinsic strain amplitude'' $h_0$ defined by
\citet{Jaranowski1998}, this indirect limit is \citep{Wette2008}
\begin{equation}
\label{indirect}
h_0 < 1.26\times10^{-24} \left( \frac{\mbox{3.30 kpc}} {D} \right) \left(
\frac{\mbox{300 yr}} {a} \right)^{1/2},
\end{equation}
where $D$ is the distance to the source and $a$ is its age.
This assumes a moment of inertia $10^{45}$~g~cm$^2$ and (spherical harmonic
$m=2$) mass quadrupole \ac{GW} emission, the usual assumptions in the \ac{GW}
literature; and also assumes that the star has spun down at least 10--20\%
since birth, the usual assumption in the pulsar literature.
For current quadrupole ($r$-mode) emission, the indirect limit on $h_0$ is
slightly higher \citep{Owen:2010ng}; but we used the mass quadrupole value.
The intrinsic strain $h_0$ is generally a factor 2--3 greater than the actual
strain amplitude response of a detector; it is defined precisely in
\citet{Jaranowski1998} and related to standard multipoles and properties of
the source in \citet{Owen:2010ng}.
It can be converted to fiducial ellipticity and $r$-mode amplitude of a
neutron star via Eqs.~(\ref{epsilon}) and~(\ref{alpha}).
In order to beat the limit~(\ref{indirect}) over as wide a frequency band as
possible, we generally used the most optimistic (lowest) age and distance
estimates from the literature, corresponding to the highest indirect limit,
with exceptions noted below.
The algorithm for that final selection is described in the next subsection.

The resulting target list and astronomical parameters are shown in
Table~\ref{t:targets}.
The individual \acp{SNR} and the provenance of the parameters used are:

\textit{G1.9+0.3}---Currently the youngest known \ac{SNR} in the galaxy
\citep{Reynolds2008}.
Nothing is visible inside the remnant, which although more than an arcminute
across is small enough to be searched with one sky position for the
integration times used here \citep{Whitbeck2006}.
Several arguments favor it being a Type~Ia \citep{Reynolds2008}, which would
leave no neutron star behind, but this is not definite and the remnant's youth
makes it an interesting target on the chance it is not Type~Ia.
We used the position of the center of the remnant from the discovery paper
\citep{Reich1984}.
The age and distance are from the ``rediscovery'' paper \citep{Reynolds2008}.

\textit{G18.9--1.1}---The position is that of the \textit{Chandra}
point source discovered by \citet{Tullmann2010}.
There is also an x-ray and radio \ac{PWN} candidate trailing back toward the
center of the \ac{SNR} \citep{Tullmann2010}.
Age and distance estimates are from \citet{Harrus2004}.

\textit{G93.3+6.9}---Also known as DA~530.
The position and age are from \citet{Jiang2007}.
No true (sub-arcsecond) \textit{Chandra} point source is seen, but since the
x-ray and radio \ac{PWN} candidate is barely detected in x-rays and the
\ac{SNR} overall is anomalously faint in x-rays, it is plausible that the
pulsar powering the nebula is weak and remains to be detected in the brightest
part of the \ac{PWN}.
The e-folding scale of X-ray intensity at the center of the \ac{PWN} candidate
is 6'', which qualifies as a point source for the \ac{GW} search.
The distance estimate is from \citet{Foster2003}.

\textit{G111.7--2.1}---Also known as Cas~A.
The point source is the prototypical \ac{CCO} and was discovered with
\textit{Chandra}'s first light \citep{Tananbaum1999}.
The position is from that reference, the distance from \citet{Reed1995}, the
age from \citet{Fesen2006}.
In this search we used 300 years rather than 330 years as in \citet{S5CasA},
reflecting the idea of using optimistic ends of ranges given in the
literature, which also corresponds to broader parameter space coverage.
There is no evidence for a \ac{PWN}, indicating that the neutron star may be
slowly spinning or (better for \ac{GW} emission) that it may have a weak
surface magnetic field.

\textit{G189.1+3.0}---Also known as IC~443.
The position is that of the \textit{Chandra} point source found by
\citet{Olbert2001}.
It is a \ac{CCO}-like object, though not at the center of the remnant, buried
in a comet-shaped x-ray, radio, and possibly $\gamma$-ray \ac{PWN}.
This object is often studied, with a wide range of distance and age estimates
in the literature.
We used \citet{Petre1988} for an optimistic age estimate.
We did not use the most optimistic distance quoted, but the assumed
association with the I~Gem cluster from \citet{Fesen1980}.

\textit{G266.2--1.2}---Also known as Vela~Jr.
The position is that of the \textit{Chandra} point source found by
\citet{Pavlov2001}.
It is a \ac{CCO} with no evidence of a \ac{PWN}.
The literature on this object also features a wide range of age and distance
estimates, enough that we performed two searches (``wide'' and ``deep'').
We used \citet{Iyudin1998} for the most optimistic age and distance, which
were used in the wide search.
The more pessimistic numbers, for the deep search, are from
\citet{Katsuda2008}.
Even more extreme numbers have been quoted in the literature, but we
restricted ourselves to those publications that contained some derivations of
the numbers.
[This was true at the time the computations were performed: As this manuscript
was about to be submitted, a manuscript with derivations of more pessimistic
numbers was made public \citep{AllenEtAl}.]

\textit{G291.0--0.1}---Also known as MSH~11$-$62.
The position and age are from the \textit{Chandra} point source discovery
paper \citep{Slane2012}.
The distance is from \citet{Moffett2001}.
The age and distance are derived in slightly inconsistent ways, but rather
than attempt to repeat the calculations we stuck to the numbers quoted in the
literature.
The point source is embedded in a powerful \ac{PWN} seen in x-rays and radio
and possibly $\gamma$ rays, although the poor Fermi-LAT spatial resolution
makes the latter identification uncertain \citep{Slane2012}.

\textit{G347.3--0.5}---\citet{Mignani2008} obtained the sub-arcsecond
position from archival \textit{Chandra} data, although the \ac{CCO} had been
identified in ASCA data earlier \citep{Slane1999}.
There is no evidence of a \ac{PWN}.
We used the distance from \citet{Cassam-Chenai2004} and the age from the
proposed identification with a possible SN~393 \citep{Wang1997}.
Although this identification may be problematic given the inferred properties
of such a supernova, other age estimates are comparable \citep{Fesen2012}.

\textit{G350.1--0.3}---Position and distance estimates are from the
discovery paper of the \textit{XMM-Newton} point source by
\citet{Gaensler2008}.
This is a \ac{CCO} candidate with no evidence of a \ac{PWN}.
The age is from \textit{Chandra} observations \citet{Lovchinsky2011}.

\subsection{Target selection and search parameters}
\label{ss:parameters}

\begin{table}
\caption{\label{t:outliers}Outliers warranting manual investigation}
\begin{center}
\begin{tabular}{lrrl}
\hline
\hline
Search & \multicolumn{2}{c}{Job min.\ and max.} & Note
\\
& \multicolumn{2}{c}{frequency (Hz)} &
\\
\hline
G18.9$-$1.1 & 192.470 & 192.477 & Pulsar 8
\\
G189.1$+$3.0 & 393.167 & 393.176 & H1 \& L1 clock noise
\\
G189.1$+$3.0 & 399.264 & 399.272 & L1 clock noise
\\
G266.2$-$1.2 wide & 441.004 & 441.212 & H1 geophone
\\
G266.2$-$1.2 wide & 1397.720 & 1397.780 & Pulsar 4
\\
G266.2$-$1.2 wide & 1408.100 & 1408.170 & H1 electronics
\\
G347.3$-$0.5 & 108.790 & 108.920 & Pulsar 3
\\
G347.3$-$0.5 & 192.448 & 192.522 & Pulsar 8
\\
G350.1$-$0.3 & 192.465 & 192.472 & Pulsar 8
\\
G350.1$-$0.3 & 192.472 & 192.479 & Pulsar 8
\\
\hline
\end{tabular}
\end{center}
Search jobs that produced non-vetoed candidates above the 95\% confidence
(5\% false alarm probability) Gaussian threshold, along with the most likely
causes.
Notes of the form ``Pulsar $N$'' refer to hardware-injected signals (see the
Appendix).
The others are described in the text.
Frequencies are shown in the solar system barycenter frame at the beginning of
each observation span.
\end{table}

The final selection of targets involved estimating \ac{GW} search
sensitivities and computing costs to determine which objects could feasibly be
searched well enough to beat the energy conservation limits on \ac{GW}
emission---see Eq.~(\ref{indirect}).
The sensitivity of each search was worked out in two iterations.

The first iteration made an optimistic sensitivity estimate using the noise
\ac{PSD} harmonically averaged over all \ac{S6} and both \ac{LIGO}
interferometers.
Writing the 95\% confidence upper limit (see Section~\ref{s:uls}) on intrinsic
strain $h_0$ as
\begin{equation}
\label{sens1}
h_0^{95\%} = \Theta \sqrt{ \frac{S_h} {T_\mathrm{data}} },
\end{equation}
where $T_\mathrm{data}$ is the total data live time, the first iteration used
a threshold factor $\Theta$ of 28 to ensure that it was too optimistic and
thus did not rule out any targets that the second iteration would find
feasible.
[The second iteration results are not sensitive to the precise $\Theta$ chosen
in the first iteration, as long as the first iteration value is slightly lower
than the true values, which are in the 30s as was seen in \citet{S5CasA} and
in the results of the second iteration.]

For a given frequency, we chose the range of first and second frequency
derivatives in the same manner as \citet{S5CasA}.
That is, we assumed a range of braking indices $n=f\ddot{f}/\dot{f}^2$ from
2--7, so that
\begin{equation}
-\frac{f} {(n_{\min}-1)a} \le \dot{f} \le -\frac{f} {(n_{\max}-1)a}
\end{equation}
at each frequency.
For each $(f,\dot{f})$ the second derivative satisfied
\begin{equation}
\frac{n_{\min} \dot{f}^2} {f} \le \ddot{f} \le \frac{n_{\max} \dot{f}^2} {f}.
\end{equation}
The physical reasoning behind these choices is explained further in
\citet{S5CasA}.
The goal is to cover observed braking indices plus a variety of predicted
\ac{GW} emission mechanisms whose relative importance may have changed over
time.
Note that the range of $\dot{f}$ does not extend up to zero.
This might seem to be an issue as it would not include ``anti-magnetars'', or
young neutron stars which are observed to spin down very slowly and hence must
have small surface magnetic fields \citep{Gotthelf2008}.
However, these are stars we would not detect anyway---any star with \ac{GW}
emission close enough to the indirect limit to be detected would have a high
spin-down due to that emission, even if it had a low surface magnetic field.

The computational cost is a function of the parameter space covered.
That functional dependence was used to choose the parameters of these searches
and will be used in planning future searches.
The product of the ranges on $f$, $\dot{f}$, and $\ddot{f}$ suggests that the
size of the parameter space and the computational cost should scale as
$f_{\max}^3 a^{-3} T_\mathrm{span}^7$ \citep{Wette2008}.
In the limit that only one value of $\ddot{f}$ is used, the range of that
parameter should be eliminated from the product, the parameter space should be
two dimensional rather than three, and the scaling should be $f_{\max}^2
a^{-1} T_\mathrm{span}^4$.
By setting up several searches with different parameters perturbed from those
of the Cas~A search, we observed that the computational cost scaled roughly as
$f_{\max}^{2.2} a^{-1.1} T_\mathrm{span}^4$.
Comparing this to \citet{Wette2008} shows that the effective dimensions of the
template banks were nearly 2 rather than 3, as confirmed by the fact that the
number of different $\ddot{f}$ values in the template banks was typically more
than one but small.

Assuming a 70\% duty factor, and the empirical scaling for computational cost
above, we determined the three unknowns $(f_{\min}, f_{\max},
T_\mathrm{span})$ by setting the sensitivity~(\ref{sens1}) equal to the
indirect limit on $h_0$~(\ref{indirect}) at both ends of the search frequency
band ($f_{\min}$ and $f_{\max}$).
The third condition to fix the three unknowns was to keep the computational
cost per search at roughly the same nominal value as \citet{S5CasA}, although
because of hardware and software improvements the total computational time was
less (see below).

The second iteration involved running the analysis pipeline on small bands to
get true template densities, the noise \ac{PSD} of the optimal data stretch
for each search, upper limits, and thus a better estimate of each $\Theta$.
For at least a 10~Hz band near each $f_{\min}$ and $f_{\max}$, we ran the
search (without looking at detection candidates) to get upper limits.
We then read off the value of $\Theta$ [from the observed upper limits and
inverting Eq.~(\ref{sens1})] at frequencies near $f_{\min}$ and $f_{\max}$.
These values were spot checked beforehand to verify that upper limits were
comparable to indirect limits.
This second iteration was good enough, considering calibration uncertainties
and other errors.
The lowest (best) values of $\Theta$ were comparable to the 31.25 predicted by
averaging the calculation of \citet{Wette2012} over declination, but in some
bands $\Theta$ could be more than 40 because of narrow noisy and/or
non-stationary bands.
In general $\Theta$ rose slightly at higher frequencies because of the
increasing density of templates (per Hz).

Table~\ref{t:targets2} lists the targets and other \ac{GW} search parameters
determined by the sensitivity algorithm.
These parameters were confirmed by several consistency checks:

For each search we checked that $\ddot{f}$ was the highest frequency
derivative needed for the resulting $T_\mathrm{span}$ using the
parameter-space metric of \citet{Whitbeck2006}.
Specifically, we computed the diagonal metric component for the third
frequency derivative and verified that the $2\mathcal{F}$ lost by neglecting
that derivative in the worst corner of parameter space searched was much less
than the 20\% template bank mismatch:
In the worst case, the Vela Jr.\ wide search, it was just under 1\%.

For each search we also checked the ``pixel size'' obtained from the metric on
the sky position parameters to verify that more than one sky position was not
needed.
The position error ellipses for a 20\% mismatch were roughly 0.8--2 arcminutes
across the minor axis for $T_\mathrm{span}$ of two weeks, and that width
scaled as the inverse of $T_\mathrm{span}$.
Most of the target positions are known to sub-arcsecond accuracy.
The location of the object in \ac{SNR}~G93.3+6.9 is known to a few arcseconds.
\ac{SNR}~G1.9+0.3 has no known object inside, but the remnant itself is barely
an arcminute across; and given the age and distance any neutron star would
have moved only a few arcseconds from the center of the remnant even at
transverse kick velocities of order 1000~km/s.
Since the integration time for that \ac{SNR} was short, the error ellipse was
several arcminutes across.

We also confirmed that the standard 1800-second \acp{SFT} do not cause
problems.
The $\mathcal{F}$-statistic code requires that signals not change more than a
frequency bin over the duration of an \ac{SFT}.
The maximum $\dot{f}$ feasible is then 1/(1800~s)$^2 \approx
3\times10^{-7}$~Hz/s.
The strongest $\dot{f}$ from orbital motion in these searches is
2~kHz$\times 10^{-4} \times 2\pi$/1~yr$\approx 4\times10^{-8}$~Hz/s, where the
$10^{-4}$ is the Earth's orbital velocity in units of $c$.
The strongest intrinsic spin-down is 2~kHz/690~yr$\approx 9\times10^{-8}$Hz/s.
(Both of these figures come from the Vela Jr.\ wide search.)

\subsection{Implementation}

\begin{figure*}
\includegraphics[width=0.47\textwidth]{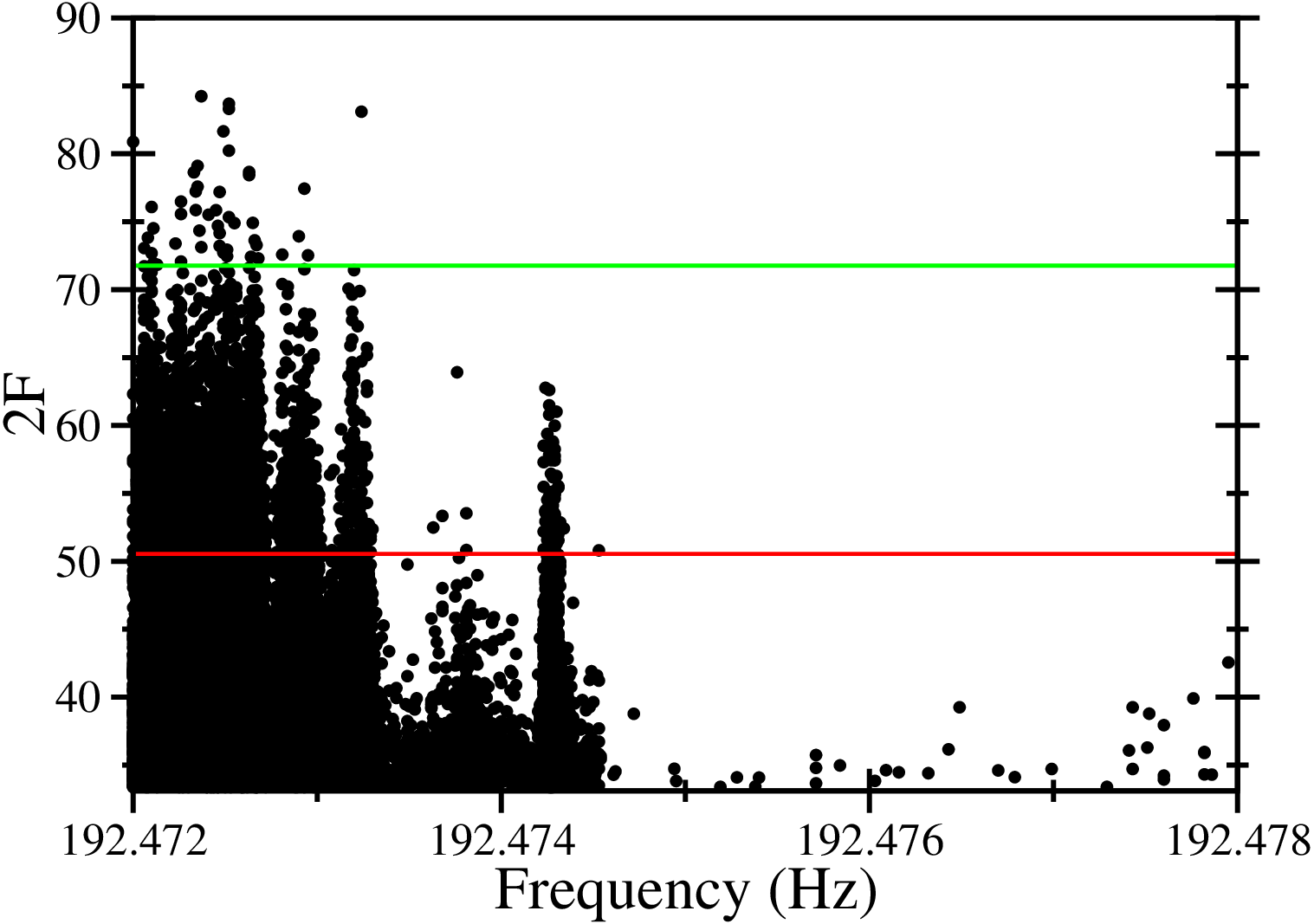}
\hspace{0.02\columnwidth}
\includegraphics[width=0.47\textwidth]{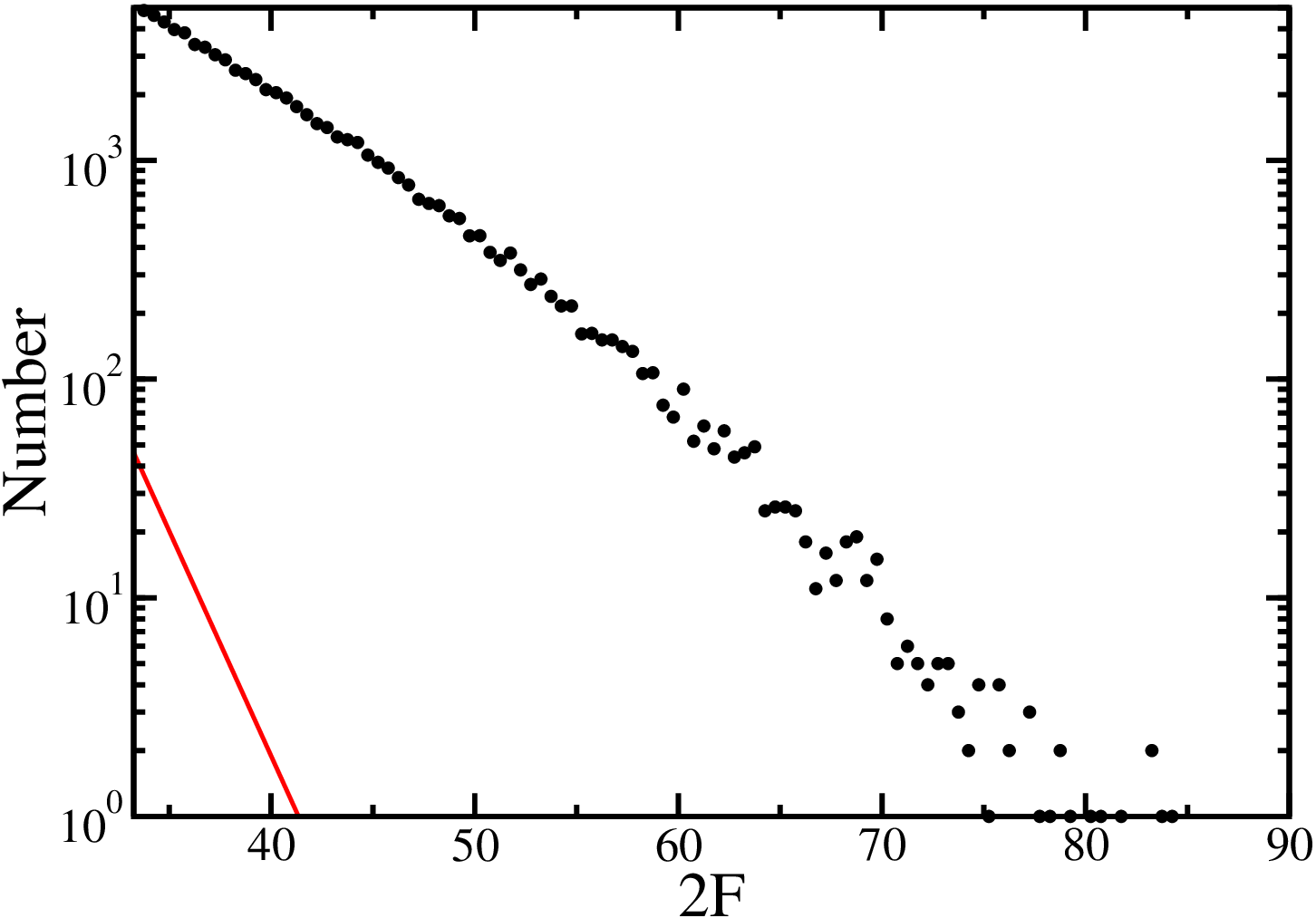}
\vspace{8ex}
\newline
\includegraphics[width=0.47\textwidth]{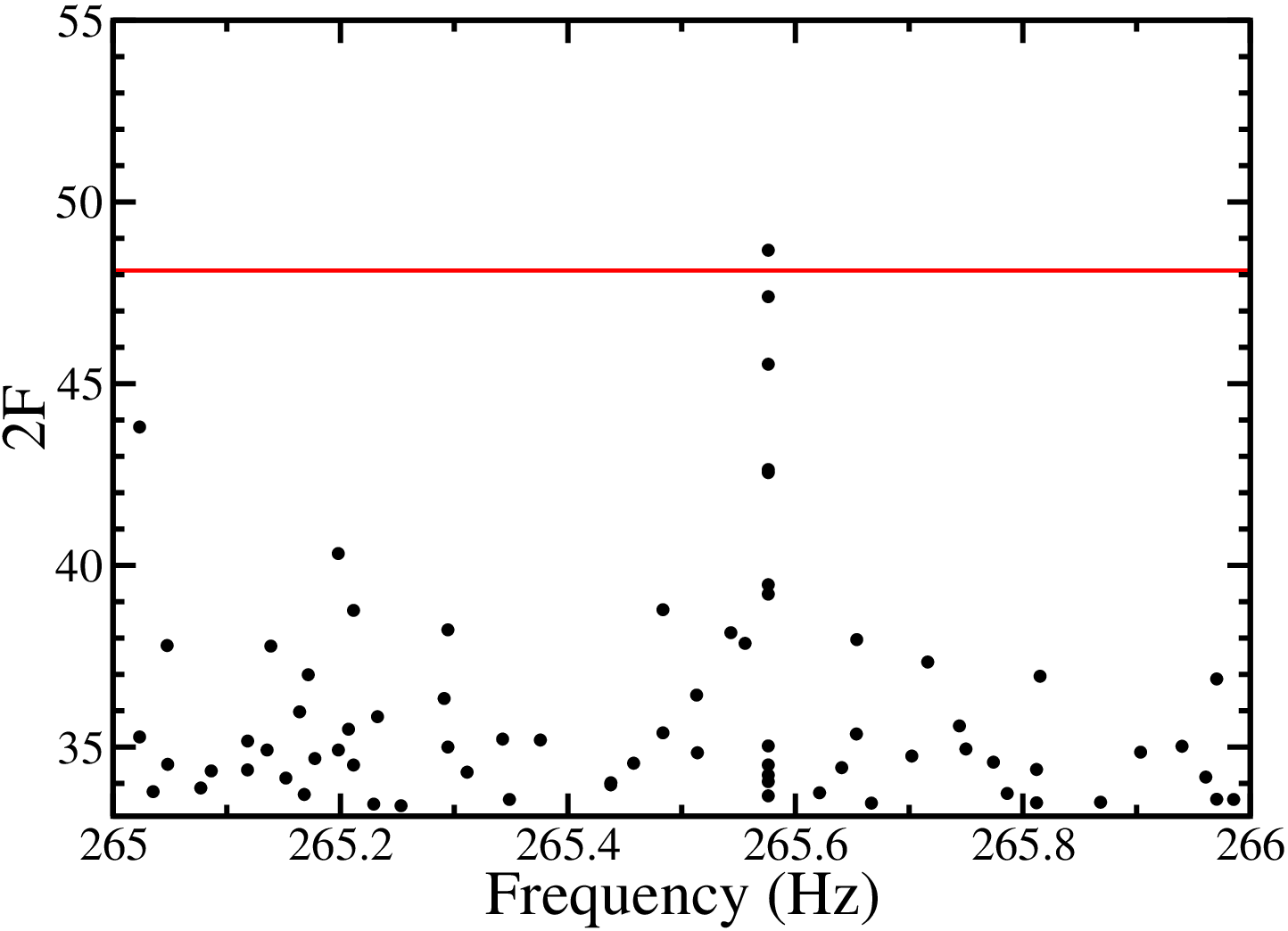}
\hspace{0.02\columnwidth}
\includegraphics[width=0.47\textwidth]{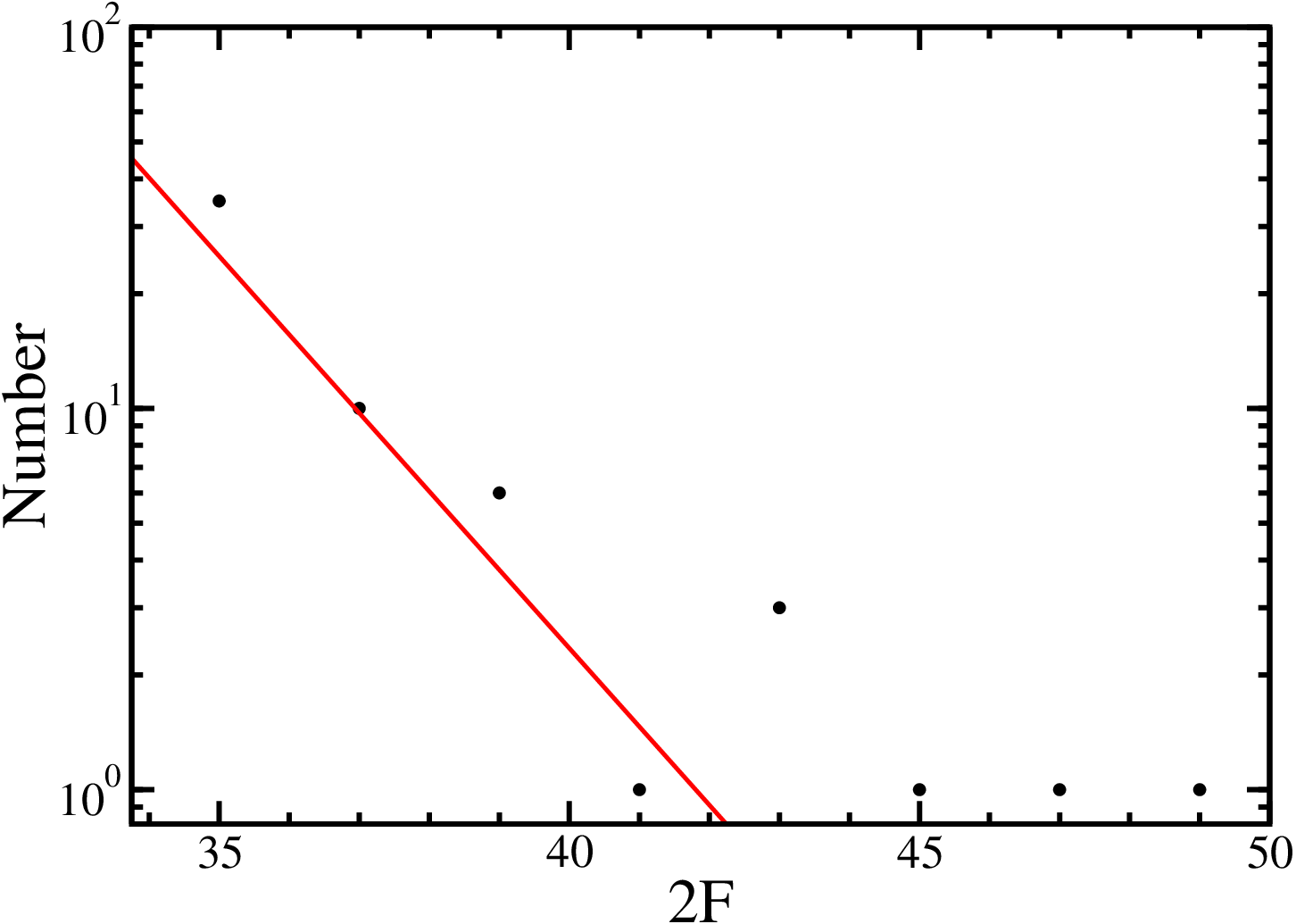}
\caption{
\label{f:inspection}
Inspection of the last outlier (top) and hardware-injected Pulsar~0 (bottom).
Top left: $2\mathcal{F}$ vs.\ frequency for the search job.
The higher line is the 95\% confidence (5\% false alarm) Gaussian threshold
for the whole search; the lower line is the same for that search job.
Top right: Histogram (tail) of $2\mathcal{F}$ for the search job.
The line is for Gaussian noise, a $\chi^2$ with four degrees of freedom.
Bottom left: $2\mathcal{F}$ vs.\ frequency for the hardware injection search
job; the line is the 95\% confidence (5\% false alarm) Gaussian threshold for
that job.
Bottom right: Histogram (tail) of $2\mathcal{F}$; the line is a $\chi^2$ with
four degrees of freedom.
}
\end{figure*}

All searches ran on the Atlas computing cluster at the Max Planck Institute
for Gravitational Physics (Albert Einstein Institute) in Hanover, Germany.
Most searches used 140\,000--150\,000 computational core-hours on Intel Xeon
3220 processors, except the Vela~Jr.\ wide search which used about 110\,000.
The costing algorithm became less accurate for that search because the
effective dimensionality of the parameter space was closer to 3 than to 2, as
the range of $\ddot{f}$ searched was more than usual.
This will need to be accounted for in future searches over wide bands and/or
short spans.
The number of matched filtering templates used in each search was about
3--12$\times10^{12}$, comparable to the $7\times10^{12}$ used in
\citet{S5CasA}.
The latter cost about 420\,000 core-hours; the factor of 3 speed-up was due
mainly to the SSE2 floating-point extensions used in the new code.

The way the search costs were split into cluster computing jobs affected the
automated vetoes described in the next subsection.
Each search was split into nominal 5-hour jobs, typically 28\,000--30\,000
jobs per search, except the Vela~Jr.\ wide search which was about 22\,000.
In order to keep the search jobs at roughly the same computational cost, the
frequency band covered by each job varied with frequency.
The Vela Jr.\ wide search had jobs covering bands from 35~mHz to nearly 2~Hz
at low frequencies, while the other searches had search job bands on the order
of a few~mHz to tens of mHz.
Each search job recorded all candidates with $2\mathcal{F}$ above about 33.4,
or 1 per million in stationary Gaussian white noise.
In bands with ``clean'' noise, typical jobs with a few times $10^8$ templates
thus recorded a few hundred candidates.
This choice of recording (which was different from the \ac{S5} search which
recorded the loudest 0.01\% of events) was needed to ease the manual
investigation of outliers surviving the automated vetoes by making sure to
record some noise at expected Gaussian levels.
Such investigation was more important than in \citet{S5CasA} because of the
``dirtier'' nature of the \ac{S6} noise and housekeeping issues associated
with excessive disk space and input/output.
The searches recorded a total of about 800~GB of candidates.

\subsection{Vetoes}

\begin{figure*}
\includegraphics[width=0.31\textwidth]{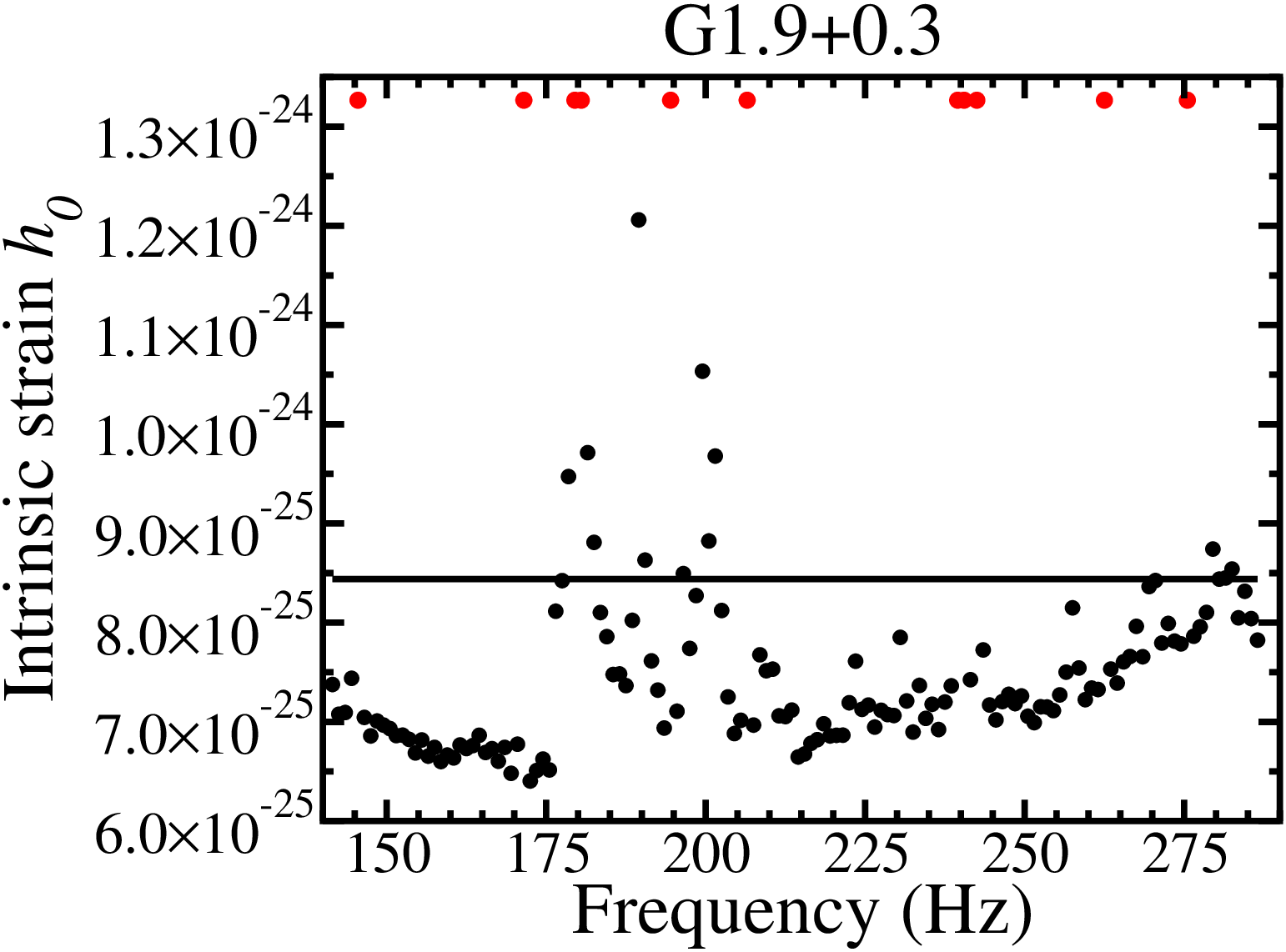}
% 11 no_uls and 135 uls
\hspace{0.02\textwidth}
\includegraphics[width=0.31\textwidth]{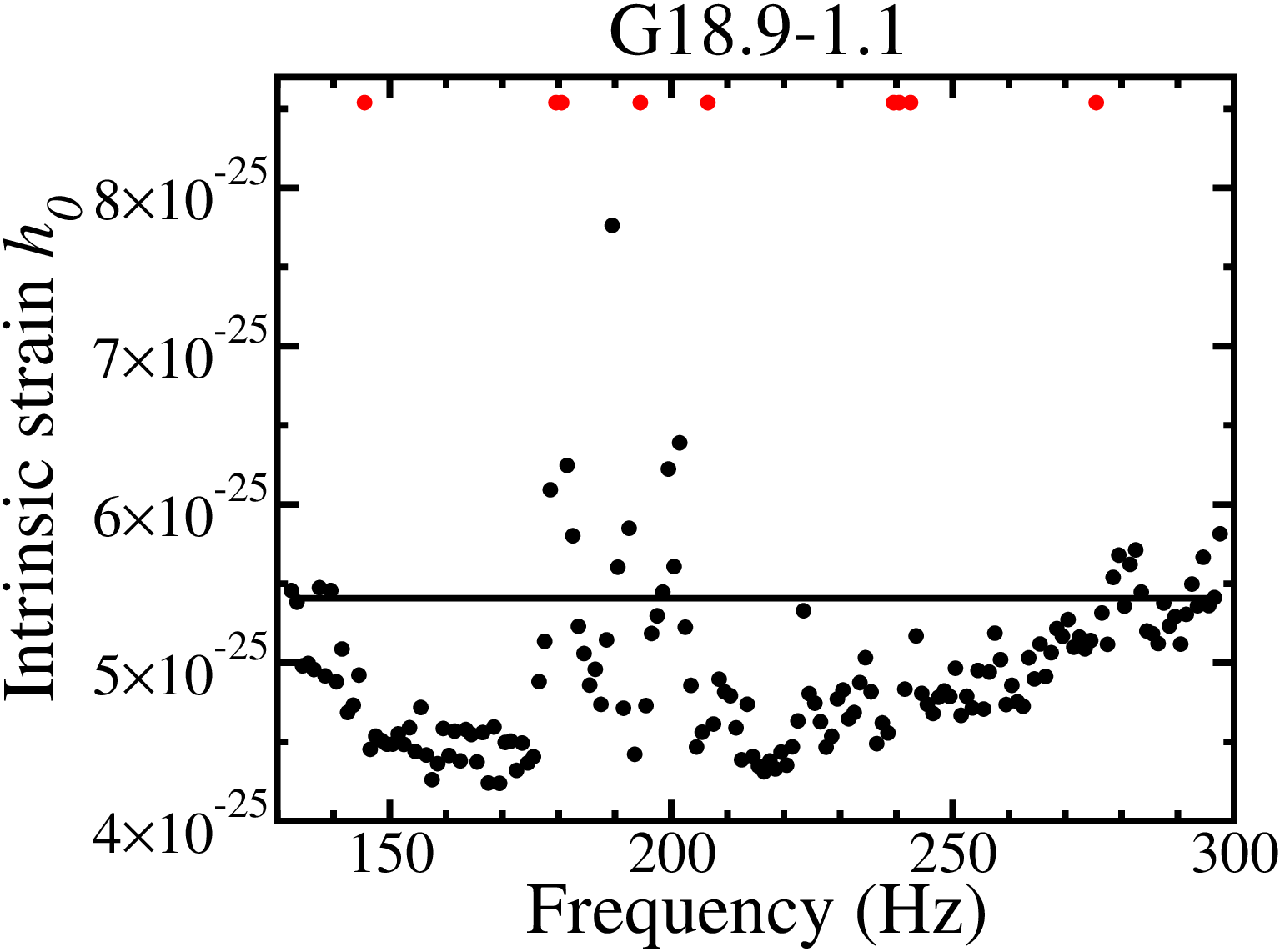}
% 9 no_uls and 157 uls
\hspace{0.02\textwidth}
\includegraphics[width=0.31\textwidth]{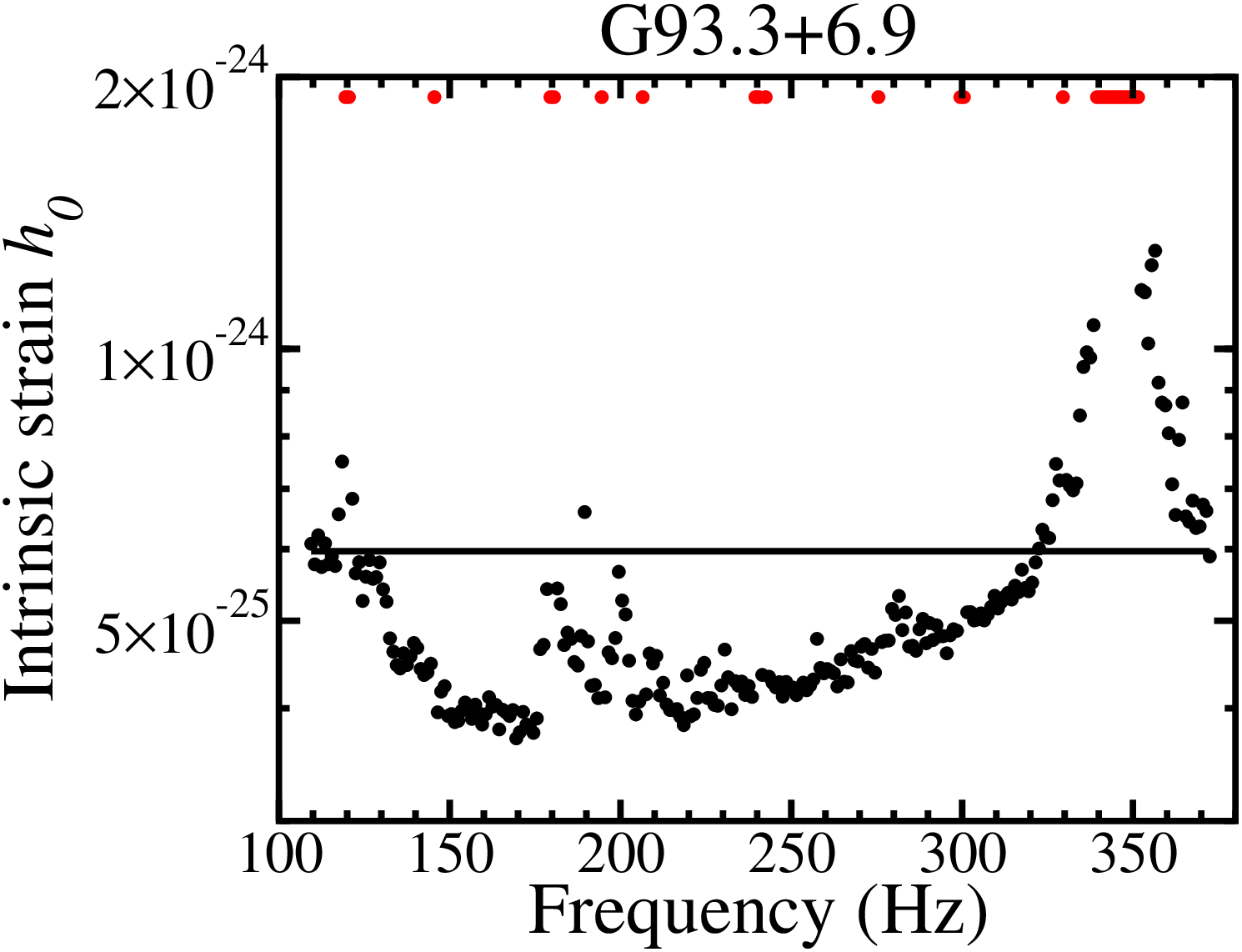}
% 27 no_uls and 237 uls
\vspace{2ex}
\newline
\includegraphics[width=0.31\textwidth]{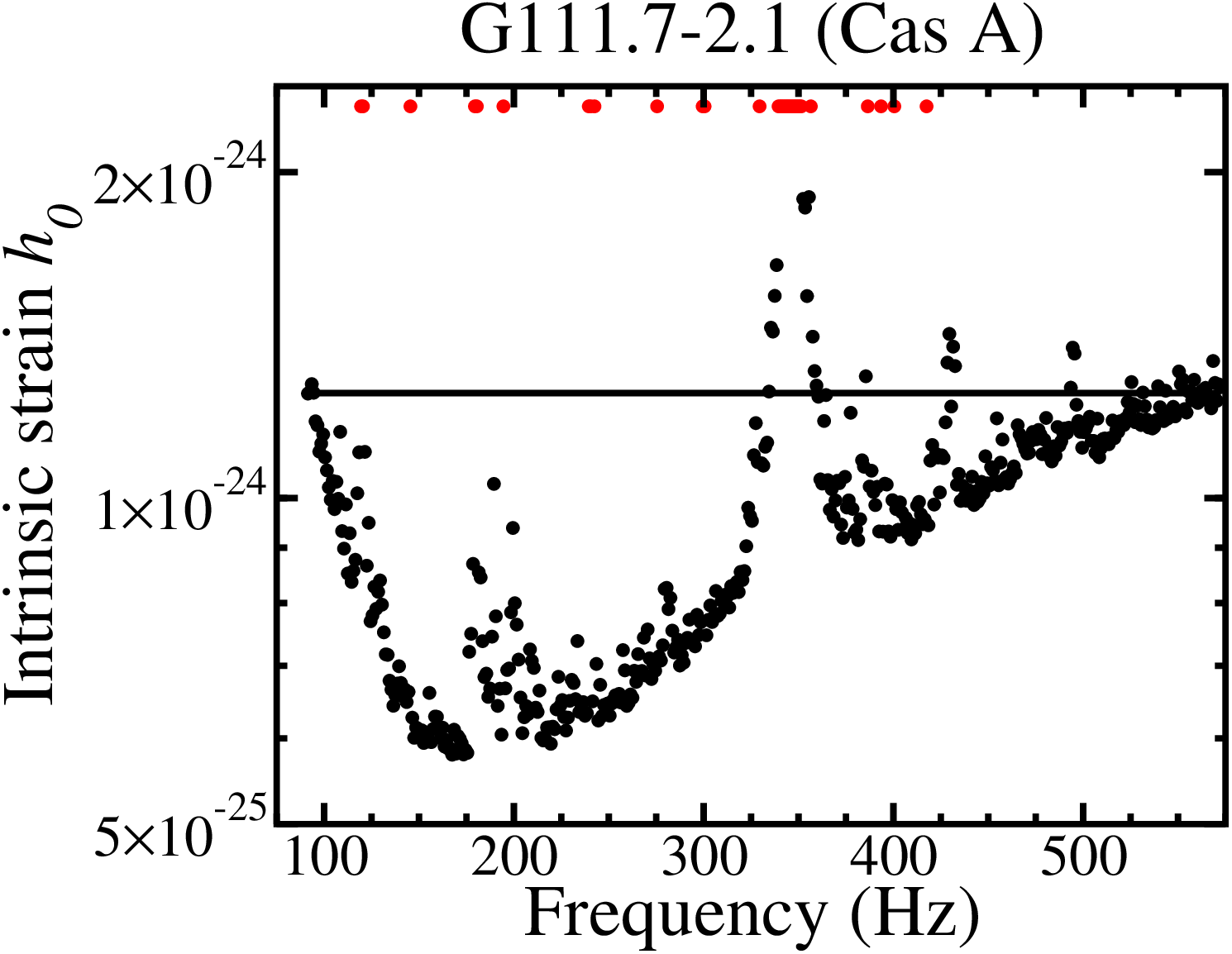}
% 31 no_uls and 451 uls
\hspace{0.02\textwidth}
\includegraphics[width=0.31\textwidth]{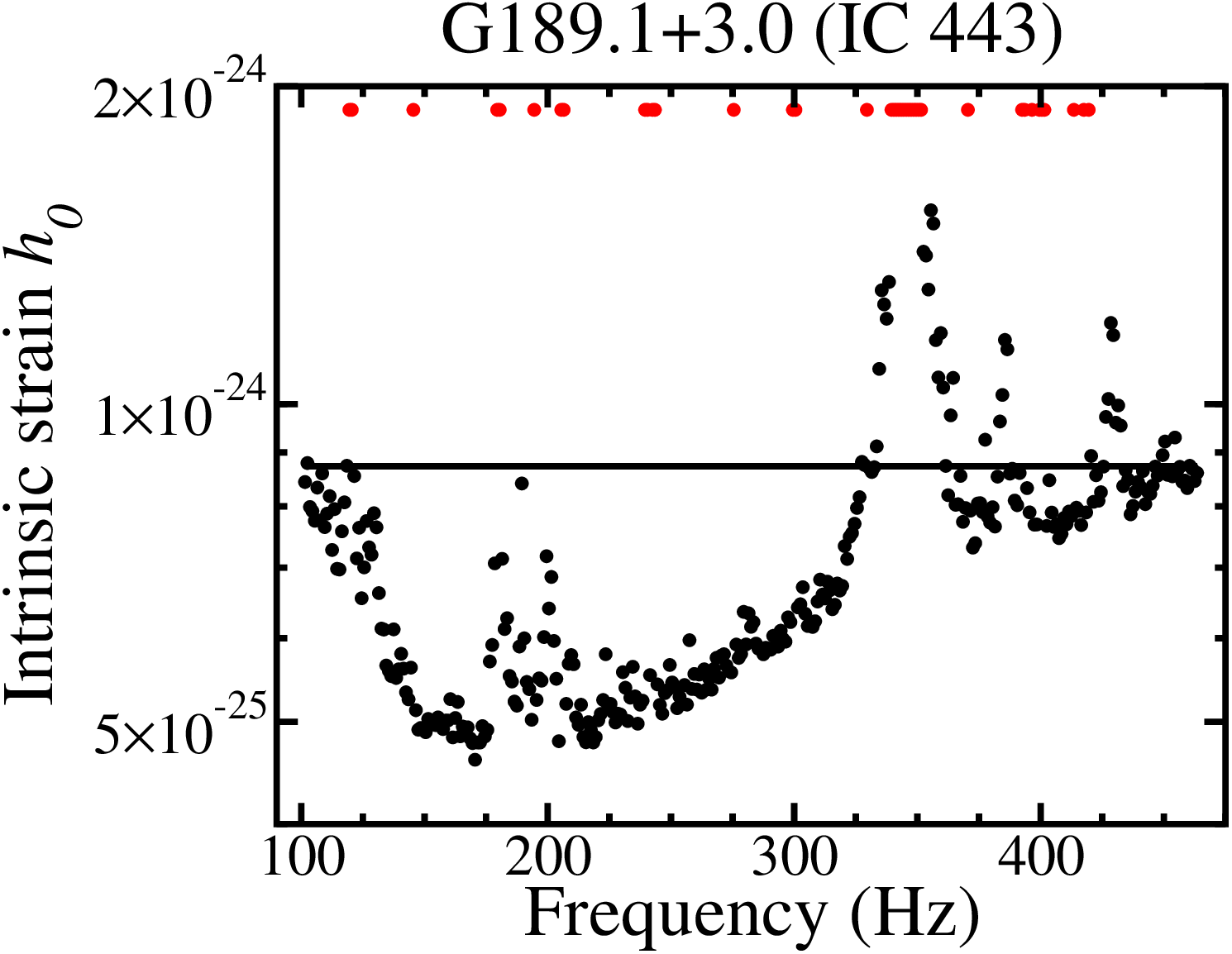}
% 39 no_uls and 324 uls
\hspace{0.02\textwidth}
\includegraphics[width=0.31\textwidth]{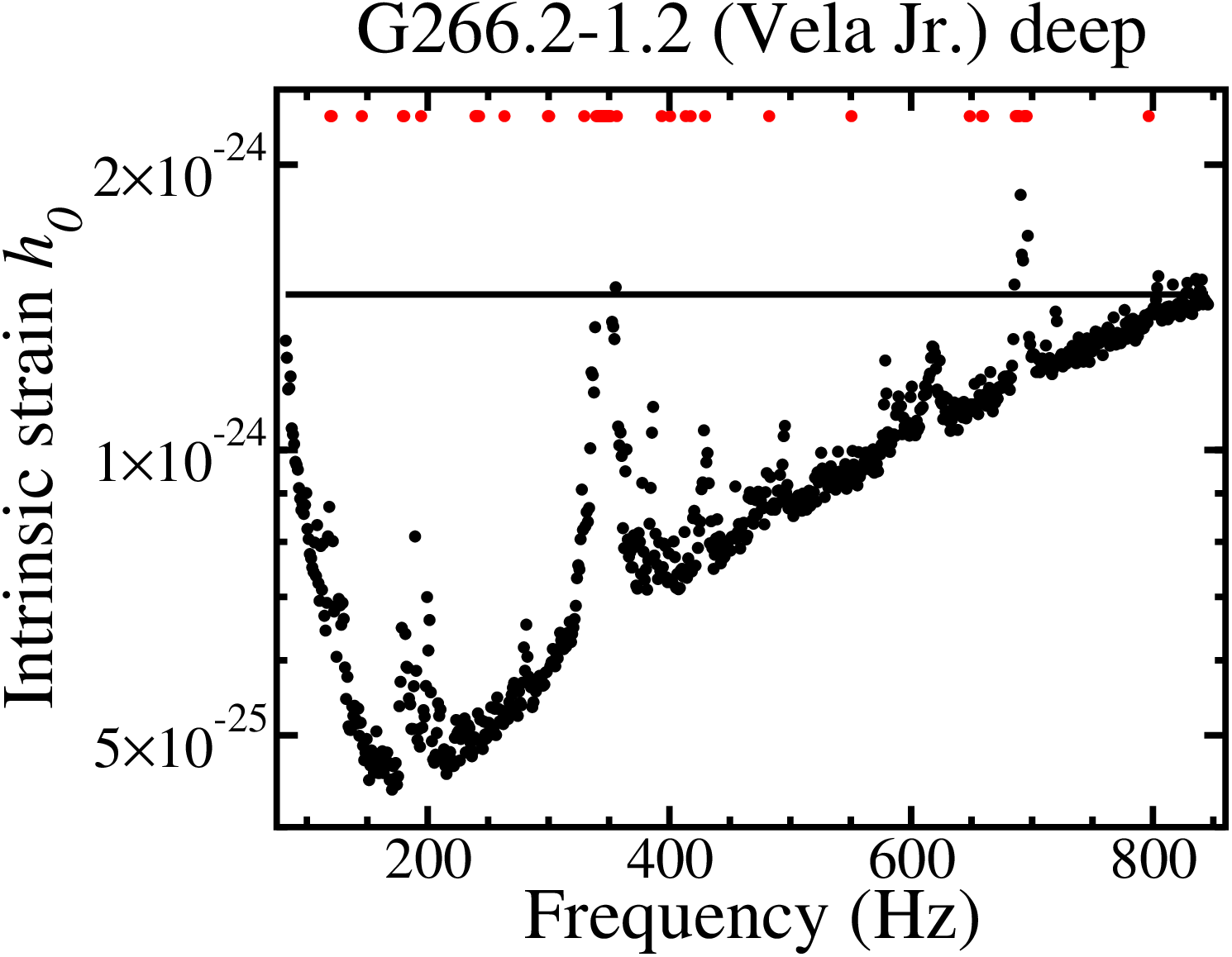}
% 45 no_uls and 719 uls
\vspace{2ex}
\newline
\includegraphics[width=0.31\textwidth]{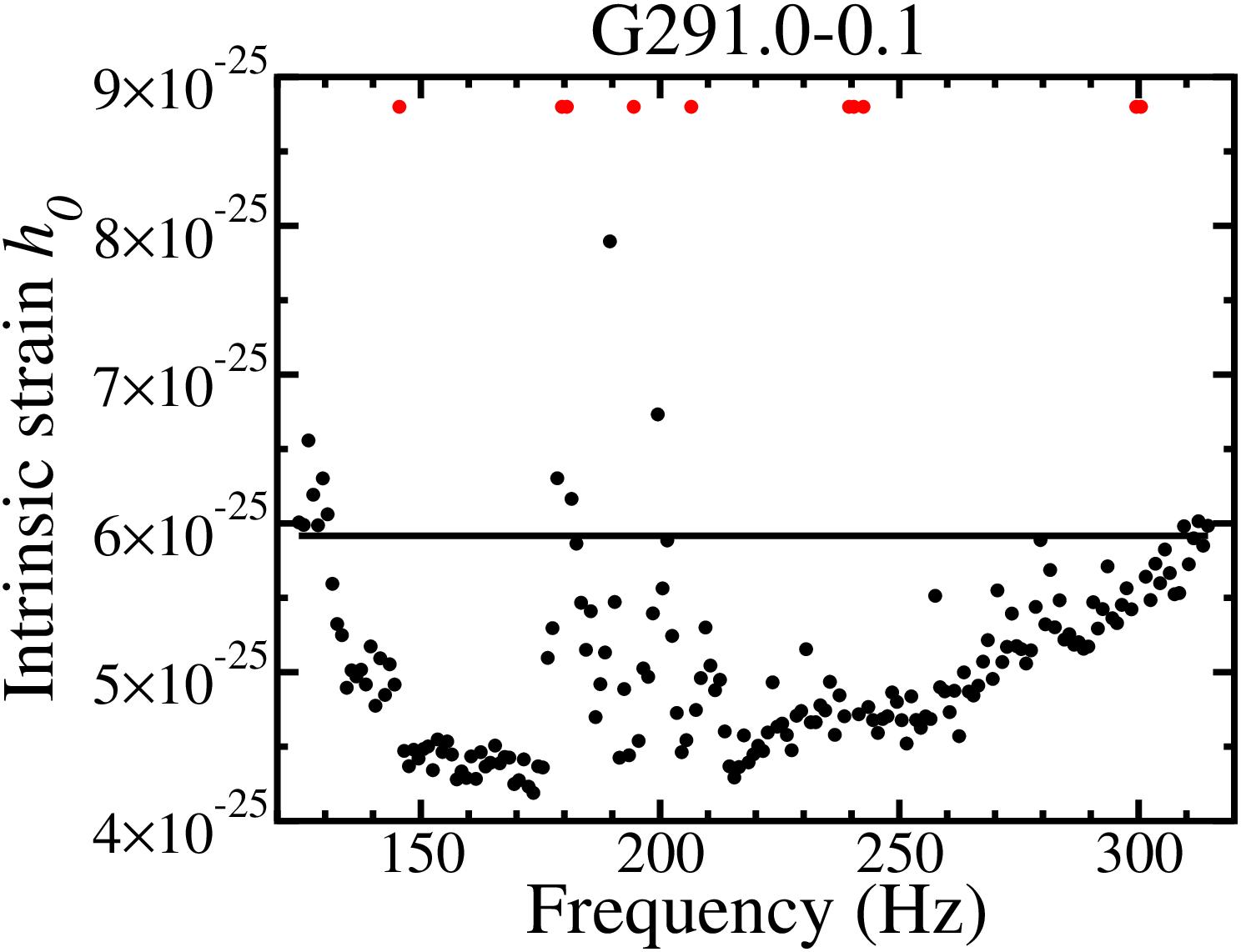}
% 10 no_uls and 181 uls
\hspace{0.02\textwidth}
\includegraphics[width=0.31\textwidth]{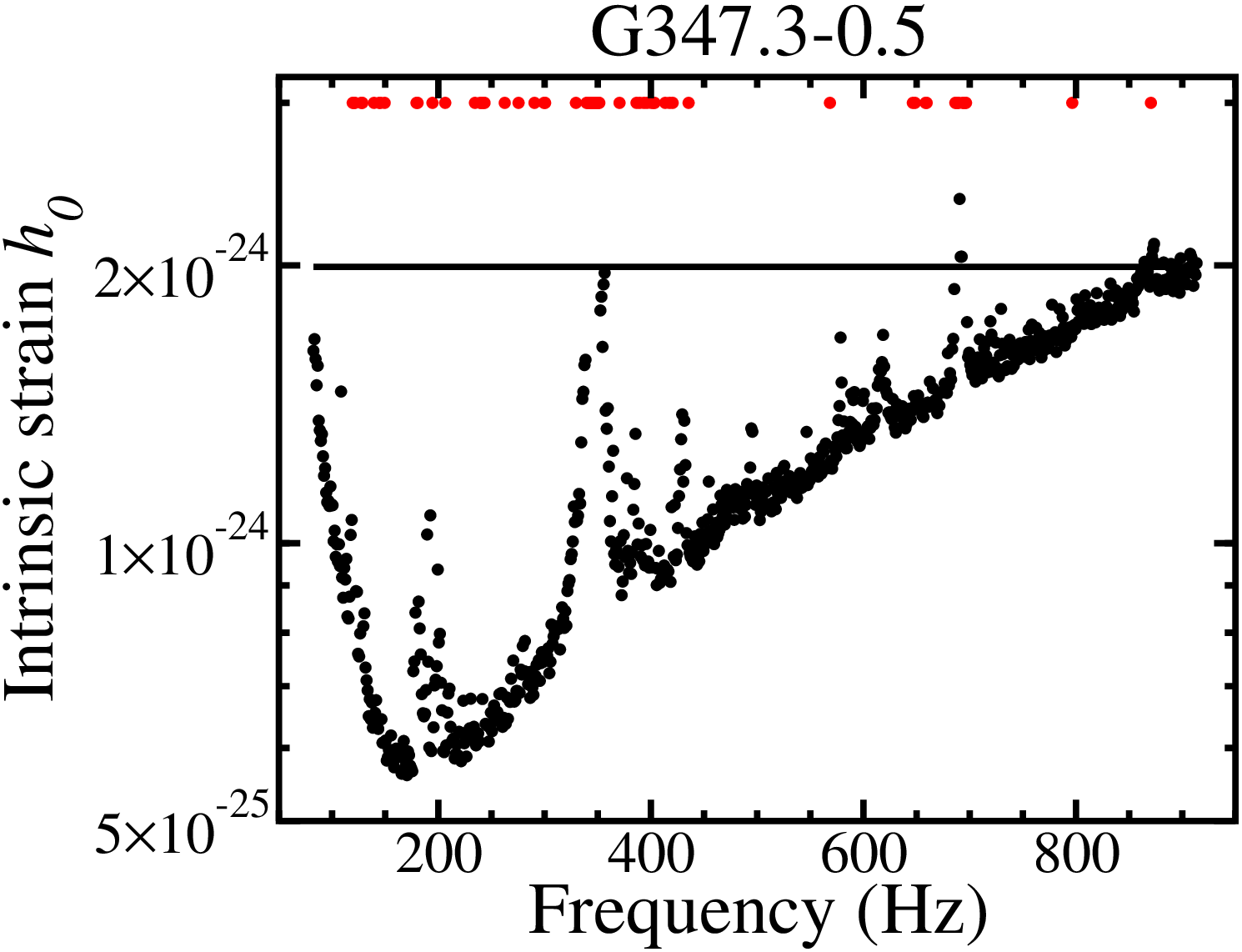}
% 64 no_uls and 768 uls
\hspace{0.02\textwidth}
\includegraphics[width=0.31\textwidth]{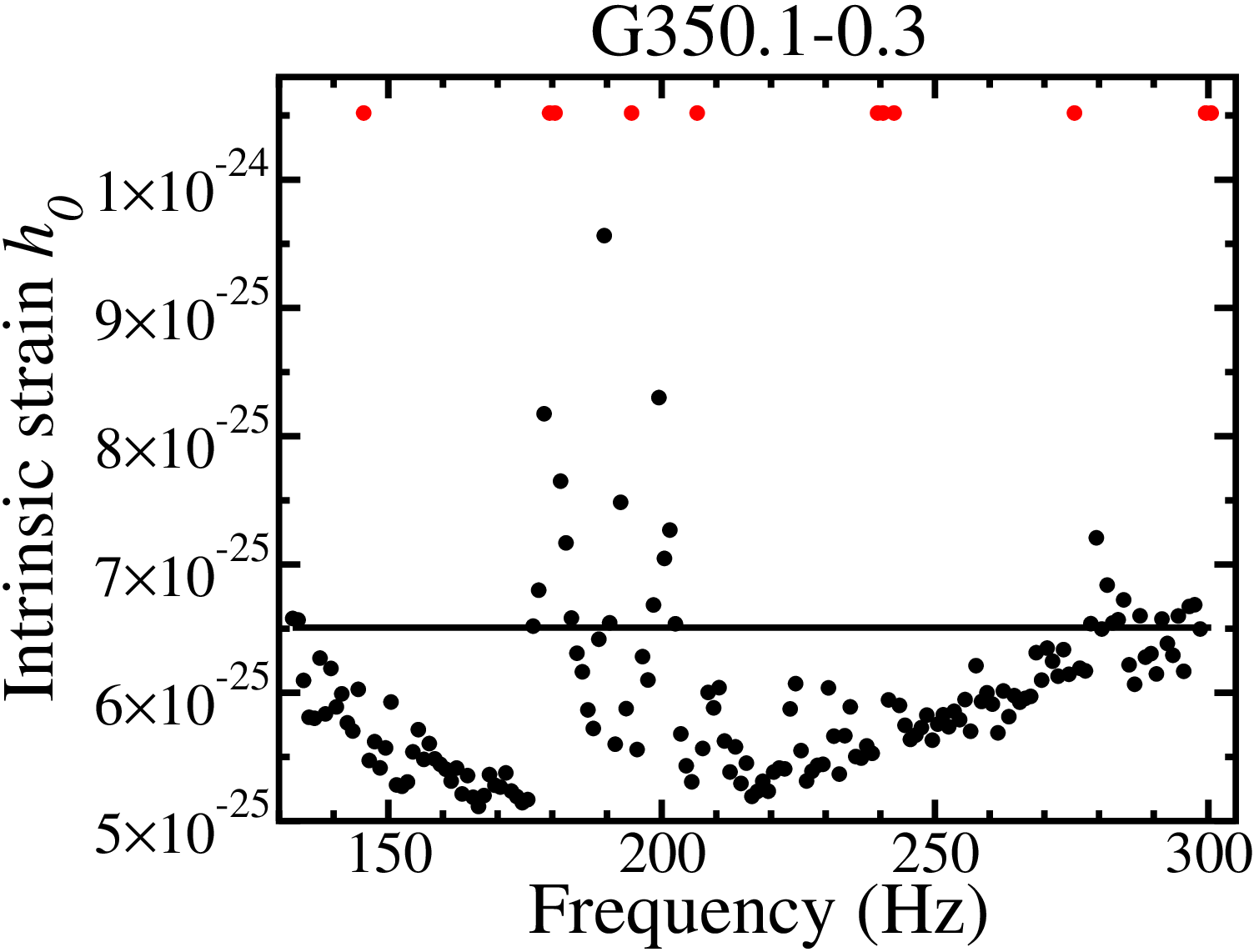}
% 11 no_uls and 158 uls
\caption{
Direct observational upper limits (95\% confidence, 5\% false dismissal) on
intrinsic strain $h_0$ are plotted as a function of frequency for all searches
except the Vela~Jr.\ wide search.
They are shown as dots (black in the on-line version), each one representing
an upper limit over a 1~Hz frequency band.
Bands where no upper limit is set (see text) are given an artificial value so
as to form a visibly distinguishable line of dots (in red in the on-line
version) near the top of each plot.
These bad bands consist of 5--10\% of the total for each search.
The solid horizontal lines are indirect limits on $h_0$ based on the ages of
and distances to the remnants.
\label{f:uls1}
}
\end{figure*}

A high value of $2\mathcal{F}$ is not enough to claim a detection, since
instrumental lines lead to non-Gaussian and/or non-stationary noise in many
narrow frequency bands.
Hence we vetoed many candidates before further investigating a few survivors.

First, we used an ``\fscan\ veto'' similar to the one used in \citet{S5CasA}.
An \fscan\ is a normalized spectrogram formed from the \acp{SFT}.
First it normalizes \acp{SFT} by scaling the power to the running median over
50 frequency bins, correcting for the bias between the finite-point running
median and the mean.
(While more complicated than simply normalizing to the mean, this procedure is
more robust to fluctuations in the time or frequency domain.)
Then the \fscan\ time-averages the normalized power in each \ac{SFT} frequency
bin.
In stationary Gaussian white noise the \fscan\ power for $N_\mathrm{SFT}$
\acp{SFT} is drawn from a $\chi^2$ distribution with $2N_\mathrm{SFT}$ degrees
of freedom scaled to unit mean (thus having a variance $N_\mathrm{SFT}$).
Therefore deviations from a $\chi^2$ indicate nonstationarity, spectral lines,
or both.

In \citet{S5CasA}, the \fscan\ veto was triggered at a threshold of 1.5 times
the expected power, which was about 11 standard deviations for H1 and 10.5 for
L1.
When triggered, it vetoed all signals overlapping a region 16 frequency bins
on either side of the central frequency (the number of terms kept in the
Dirichlet kernel) since those could be contaminated as well.
Since the SSE2 code used here kept only 8 terms, we changed the window to 8
frequency bins.

In the present searches we also changed the threshold of the \fscan\ veto
because we found that the \ac{S5} threshold was too lenient: \ac{S6} data had
many more instrumental noise artifacts.
Since the highest number of \ac{SFT} frequency bins (in the Vela~Jr.\ wide
search) was about $4\times10^6$, an \fscan\ power threshold of six standard
deviations above the mean and five below would be unlikely to veto any
Gaussian noise.
We increased the \ac{S6} threshold further to $\pm7$ standard deviations to
allow for a roughly 3\% bias (at most one standard deviation for these
searches) observed in the \fscan\ power due to the effect of estimating the
\ac{PSD} with a running median over a finite number of bins \citep{PrixFStat}.

The second veto was based on the $\mathcal{F}$-statistic consistency veto
introduced in \citet{S5Einstein2}, which uses the fact that an astrophysical
signal should have a higher joint value of $2\mathcal{F}$ (combining data from
the two interferometers) than in either interferometer alone.
Recorded candidates that violate this inequality were vetoed.
This is a simpler and more lenient version of the more recent line veto
\citep{Keitel2014}.
In clean noise bands we found that it vetoed less than 1\% of the candidates
recorded.

We extended the consistency veto to limited frequency bands as follows:
For each search job's frequency band (minus any \fscan\ vetoed bands), if the
number of candidates vetoed for consistency was greater than the number of
templates not vetoed, the entire search job was vetoed as being contaminated
by a broad feature in one interferometer.
Since we kept candidates at the 1 per million level for Gaussian noise, search
jobs in clean noise bands recorded hundreds of templates, and hence this veto
was only triggered if the number of consistency-vetoed candidates was about
two orders of magnitude greater than usual.

The combination of these vetoes, although each was fairly lenient, greatly
reduced the number of candidates surviving for human inspection.
The vetoes also proved to be safe, in the sense that they were not triggered
by the hardware-injected signals, with the exception of a few injections that
were so loud that they distorted the data \ac{PSD} and made it nonstationary
(i.e.\ triggered the \fscan\ veto).
It was easy to check that no astrophysical signals were vetoed this way by
verifying that the small number of bands vetoed in both interferometers were
due to the loud hardware-injected signals described in the Appendix or to
known instrumental artifacts.
The total frequency band vetoed was just over 1\% of the frequency band
searched, for all searches.
We also checked with a full pipeline run of several hundred software
injections and confirmed that, for $2\mathcal{F}$ less than about 230, about
1\% went undetected due to vetoes.

\subsection{Detection criteria and results}

\begin{figure}
\includegraphics[width=0.47\textwidth]{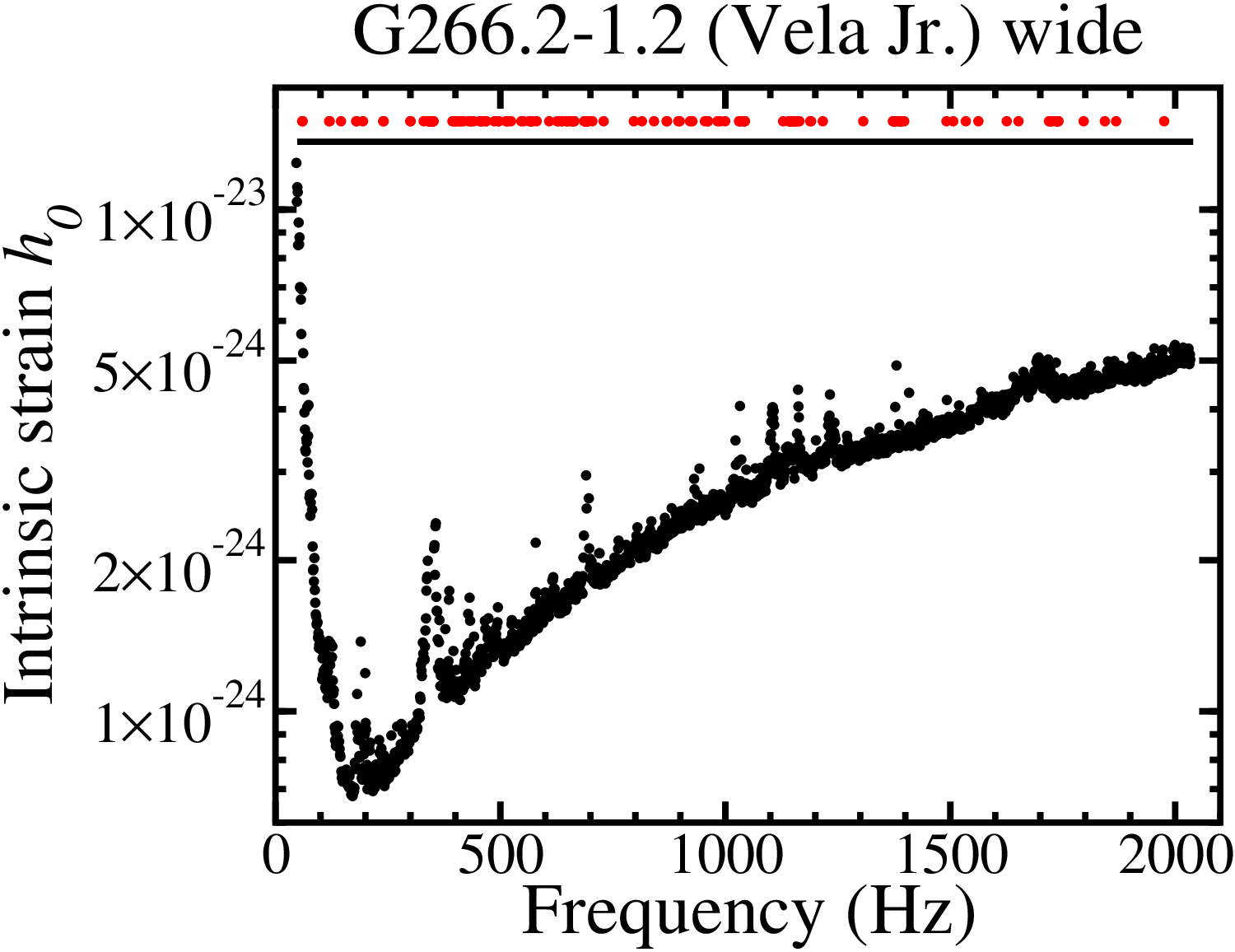}
% 156 no_uls and 1832 uls
\vspace{2ex}
\newline
\includegraphics[width=0.47\textwidth]{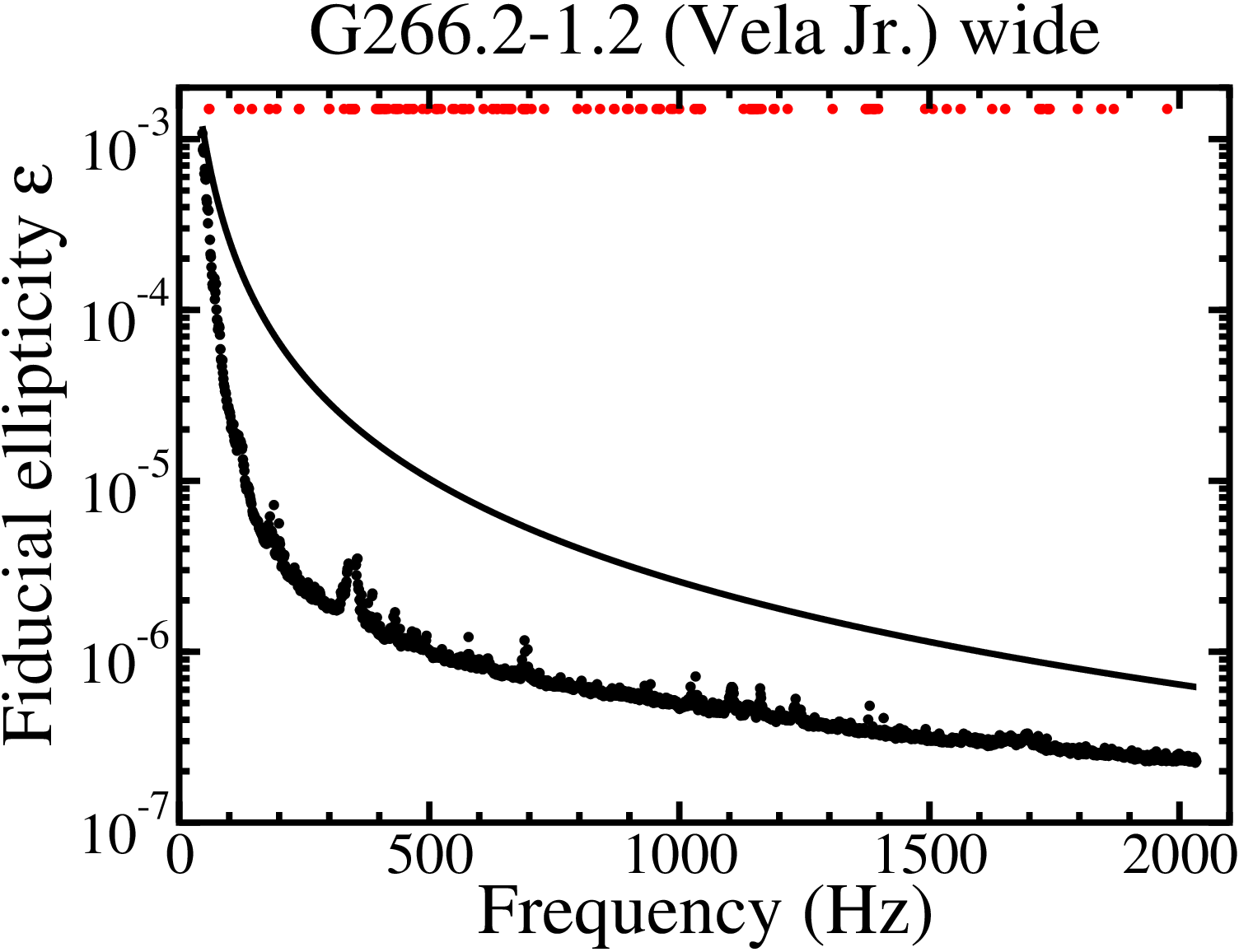}
\vspace{2ex}
\newline
\includegraphics[width=0.47\textwidth]{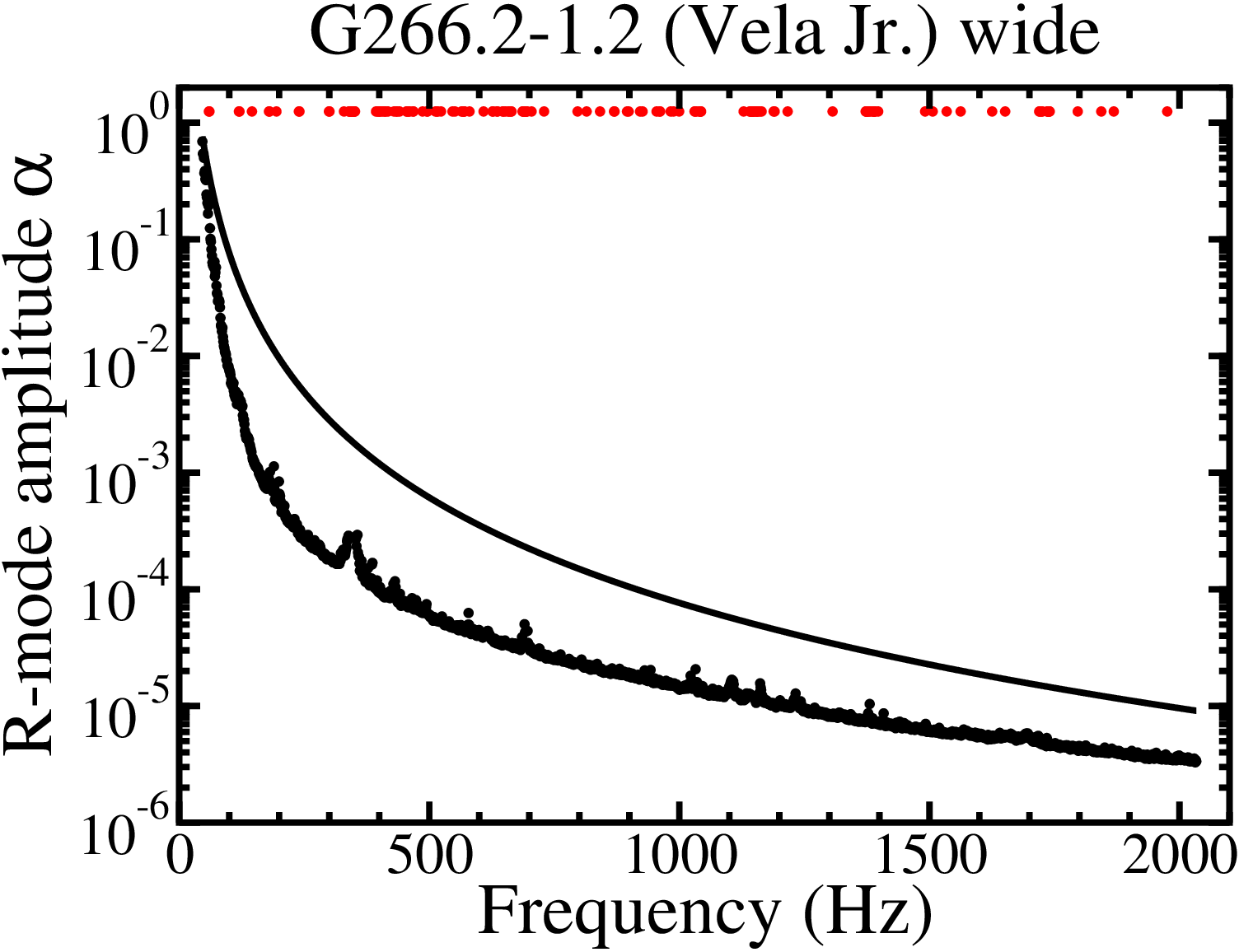}
\caption{
The top plot is the analog of Fig.~\protect\ref{f:uls1} for the Vela Jr.\ wide
search.
The middle and bottom plots are the corresponding upper limits on fiducial
ellipticity and $r$-mode amplitude.
\label{f:uls2}
}
\end{figure}

For each search, we computed the $2\mathcal{F}$ value corresponding to a 5\%
false alarm probability assuming Gaussian noise, and gave a further look to
search jobs with nonvetoed candidates passing this threshold.
Because of potential correlations between templates, we checked for an
effective number of independent templates $N_\mathrm{eff}$.
The distribution of loudest nonvetoed event per search job for each target was
nearly Gaussian.
Therefore we determined $N_\mathrm{eff}$ by minimizing the Kolmogorov-Smirnov
distance between the observed and expected cumulative distributions.
For all searches this produced $N_\mathrm{eff}$ roughly 90\% of the true
number of templates and resulted in a further-look threshold of $2\mathcal{F}
\approx $71--73.

The search jobs that produced outliers surviving the automatic vetoes and thus
warranting manual investigation are listed in Table~\ref{t:outliers}.
For all investigations it sufficed to make two plots of the results of the
search job, demonstrated in Fig.~\ref{f:inspection} for the last outlier in
Table~\ref{t:outliers} (top panels) and the first (and barely detected in 10
days' integration) hardware injection, ``Pulsar~0'' (bottom panels, see the
Appendix for more on the hardware injections).

Examples of the first plot, of $2\mathcal{F}$ vs.\ frequency, are shown in the
left-hand panels of Fig.~\ref{f:inspection}.
Injected signals showed up as near-$\delta$-functions in this plot, as in the
bottom left panel of Fig.~\ref{f:inspection}, while noise outliers had broader
structures as in the top left panel.
In most cases the outliers are clearly leaking past the edges of a vetoed
band.
Most of the outliers were near those hardware-injected signals that were loud
enough to trigger the \fscan\ veto.

The second plot used in each investigation was a histogram of the probability
density function of the recorded candidates, exemplified in the right-hand
panels of Fig.~\ref{f:inspection}.
All jobs with outliers surviving the veto process clearly showed the tail of a
$\chi^2$ distribution with the wrong normalization, as in the top right panel,
indicating that the estimator of the noise \ac{PSD} was off because of a
narrow spectral feature or nonstationarity.
Injected signals in clean data showed a correctly normalized $\chi^2$ tail
with a relatively small number of outliers extending to high $2\mathcal{F}$
values, which was visibly distinguishable from the candidates caused by noisy
data, as can be seen in the bottom right panel.

We also tracked down the instrumental sources of the outliers in
Table~\ref{t:outliers}.
(This was done after the outliers had already been dismissed by the
inspections above, and was directed toward improving future searches rather
than adding confidence to the results of this one.)
In all cases the search jobs producing outliers were adjacent in frequency
to \fscan\ vetoed bands or consistency-vetoed search jobs, and the outliers
were apparently produced by strong lines (including some very strong hardware
injections) leaking past the vetoes (which were fairly lenient).
Six of the outliers were associated with strong hardware injections, which
appeared as broad spectral features rather than $\delta$-functions due to
residual Doppler modulation (since their sky positions did not match the
positions being searched).
Of the other outliers, the first two were associated with digital clock noise
lines in both interferometers which drifted around bands of a few Hz.
In the former outlier, the lines happened to coincide at the time of the
observation; the latter outlier was just contributed by L1.
In addition, there was an outlier associated with a 441~Hz calibration signal
in a geophone prefilter in H1.
The last non-injection outlier was part of a very stable and wide-ranging
structure with dozens of sidebands seen in H1, identified also as digital
electronic noise.

\section{Upper limits}
\label{s:uls}

\subsection{Methods}

\begin{table*}
\begin{center}
\caption{
\label{t:limits}
Upper limit summary
}
\begin{tabular}{lcccccc}
\hline
\hline
Search            & Indirect $h_0$      & Direct $h_0$        & \multicolumn{2}{c}{Direct $\epsilon$} & \multicolumn{2}{c}{Direct $\alpha$}                          \\
                  &                     & lowest (best)       & at $f_{\min}$                         & at $f_{\max}$      & at $f_{\min}$      & at $f_{\max}$      \\
\tableline
G1.9$+$0.3        & $8.4\times10^{-25}$ & $6.4\times10^{-25}$ & $2.9\times10^{-4}$                    & $7.6\times10^{-5}$ & $6.2\times10^{-2}$ & $7.9\times10^{-3}$ \\
G18.9$-$1.1       & $5.4\times10^{-25}$ & $4.2\times10^{-25}$ & $5.9\times10^{-5}$                    & $1.2\times10^{-5}$ & $1.3\times10^{-2}$ & $1.2\times10^{-3}$ \\
G93.3$+$6.9       & $6.0\times10^{-25}$ & $3.7\times10^{-25}$ & $8.1\times10^{-5}$                    & $6.8\times10^{-6}$ & $2.2\times10^{-2}$ & $5.4\times10^{-4}$ \\
G111.7$-$2.1      & $1.3\times10^{-24}$ & $5.8\times10^{-25}$ & $4.6\times10^{-4}$                    & $1.2\times10^{-5}$ & $1.5\times10^{-1}$ & $6.2\times10^{-4}$ \\
G189.1$+$3.0      & $8.7\times10^{-25}$ & $4.6\times10^{-25}$ & $1.2\times10^{-4}$                    & $5.7\times10^{-6}$ & $3.4\times10^{-2}$ & $3.6\times10^{-4}$ \\
G266.2$-$1.2 wide & $1.4\times10^{-23}$ & $6.8\times10^{-25}$ & $1.1\times10^{-3}$                    & $2.3\times10^{-7}$ & $6.9\times10^{-1}$ & $3.3\times10^{-6}$ \\
G266.2$-$1.2 deep & $1.5\times10^{-24}$ & $4.4\times10^{-25}$ & $1.4\times10^{-4}$                    & $1.4\times10^{-6}$ & $4.9\times10^{-2}$ & $4.9\times10^{-5}$ \\
G291.0$-$0.1      & $5.9\times10^{-25}$ & $4.2\times10^{-25}$ & $1.3\times10^{-4}$                    & $2.0\times10^{-5}$ & $3.1\times10^{-2}$ & $1.9\times10^{-3}$ \\
G347.3$-$0.5      & $2.0\times10^{-24}$ & $5.6\times10^{-25}$ & $2.0\times10^{-4}$                    & $2.0\times10^{-6}$ & $7.3\times10^{-2}$ & $6.6\times10^{-5}$ \\
G350.1$-$0.3      & $6.5\times10^{-25}$ & $5.1\times10^{-25}$ & $1.6\times10^{-4}$                    & $3.1\times10^{-5}$ & $3.6\times10^{-2}$ & $3.1\times10^{-3}$ \\
\hline
\end{tabular}
\end{center}

Here we summarize the range of upper limits set in these searches.
The line for the G266.2$-$1.2 wide search summarizes the plots in
Fig.~\protect\ref{f:uls2} and the ``Direct $h_0$'' column summarizes the plots
in Fig.~\protect\ref{f:uls1}.
The remaining elements summarize similar results for the remaining searches
which were not plotted.
The best (lowest) upper limits on $h_0$ were set near 170~Hz for all searches,
and the corresponding limits on $\alpha$ and $\epsilon$ were near the
$f_{\max}$ of each search.
\end{table*}

The method for setting upper limits was essentially the same as in
\citet{S5CasA}.
We divided each search into 1~Hz bands.
For each of these upper limit bands, we recorded the loudest $2\mathcal{F}$
which passed the automated vetoes.
We then estimated the intrinsic strain $h_0$ at which 95\% of signals would be
found, if drawn from a population with random parameters other than $h_0$,
with a louder value than the loudest $2\mathcal{F}$ actually recorded for that
upper limit band.
That is, we set a 5\% false dismissal rate over a population of neutron stars
randomly oriented and uniformly distributed over the 1~Hz band, with the
loudest observed $2\mathcal{F}$ in that band setting the false alarm rate.

This 95\% confidence limit was first estimated for each upper limit band with
a combination of analytical and computationally cheap Monte Carlo methods.
Then, in the more computationally intensive step (in some cases 20--30\% of
the cost of the original search), we software-injected 6\,000 signals into the
band at that $h_0$ to test that the confidence level was truly 95\%.
The frequencies of these software injections were randomly chosen within the
band, and the polarization and inclination angles were chosen randomly.
The upper limit injection runs have some safety margin built in, and in fact
the confidence level was typically 96--97\%.
For a few upper limit bands---less than 1\% of the total for each
search---this test showed that the confidence level was actually lower than
95\%.
These typically corresponded to bands known to contain significant numbers of
instrumental lines, and rather than iterate the computationally expensive
procedure we chose not to present upper limits for these bands.

\subsection{Results}
\label{ss:results}

The resulting upper limits on $h_0$, in 1~Hz bands, are plotted in
Figs.~\ref{f:uls1} and~\ref{f:uls2}.
They closely follow the shape of the joint noise \ac{PSD}, although with an
overall scale factor and slight shape distortions.
The best (lowest) upper limits on $h_0$ generally occur for each search around
170~Hz, where the noise \ac{PSD} is lowest.
Several searches achieved upper limits on $h_0$ of about $4\times10^{-25}$ in
that band, as can be seen in Table~\ref{t:limits} (which also includes the
indirect limits from energy conservation).
Table~\ref{t:long} lists data for our observational upper limits on $h_0$ for
all searches, i.e.\ the black points in Fig.~\ref{f:uls1} and the top panel of
Fig.~\ref{f:uls2}, in machine-readable form.

In all these plots, the main set of points does not include bands where more
than 5\% of the 1~Hz upper limit band is vetoed or where the injection-checked
false dismissal rate was more than 5\%.
Most of these frequencies correspond to known instrumental disturbances, such
as calibration lines or clock noise.
We also removed 2~Hz bands centered on the electrical mains frequency of 60~Hz
and its harmonics up to 300~Hz, as well as the band 339--352~Hz which is full
of the extremely strong ``violin modes'' of the test mass suspension system.
While a few upper limit bands containing these lines did pass the false
dismissal and vetoed-band tests, the upper limits were much higher (weaker) on
account of the increased noise; and upper limits on bands where the noise
\ac{PSD} varies greatly within the band are not so informative.
Hence all these bad bands are removed from the main set of points, but are
plotted near the top of each plot (in red on-line, at a constant $h_0$ in each
plot) so as to give an idea of their numbers (5--10\% of the total for each
search) and locations (clustered around suspension violin modes, etc).

The strain upper limits can be converted to upper limits on the fiducial
ellipticity $\epsilon = |I_{xx} - I_{yy}| /I_{zz}$  of each neutron star using
\citep[e.g.][]{Wette2008}
\begin{equation}
\label{epsilon}
\epsilon = 9.5\times10^{-5} \left( \frac{h_0} {10^{-24}} \right)
\left( \frac{D} {\mbox{1 kpc}} \right) \left( \frac{\mbox{100 Hz}} {f}
\right)^2,
\end{equation}
assuming a fiducial value of $I_{zz} = 10^{45}$~g\,cm$^2$.
We used this equation to convert both the energy-conservation limit and the
direct 95\% confidence limits obtained here.
The results are plotted in the middle panel of Fig.~\ref{f:uls2} for the Vela
Jr.\ wide search.
This and the similar plots for the other searches are all tilted, curved
versions of the plot for $h_0$, and therefore we display only this one as an
example.
For all of the searches we summarize the ranges of ellipticity upper limits in
Table~\ref{t:limits}.

Note that this fiducial ellipticity is really a dimensionless version of the
(spherical harmonic $m=2$ part of the) mass quadrupole moment, not the true
shape of the star.
Conversion factors to these other quantities can be found in
\citet{Owen:2010ng} and \citet{Johnson-McDaniel2013}, respectively.
The quantity truly inferred from the measurement of $h_0$ (and the measured
frequency and assumed distance) is a component of the mass quadrupole.
The conversion factor to ellipticity can have uncertainties of a factor 5 or
more \citep{Johnson-McDaniel2013} depending on the neutron star mass, which
has an observed range of about a factor 2, and the equation of state, which is
significantly uncertain.

Strain upper limits can also be converted to limits on the $r$-mode amplitude
$\alpha$ \citep{Lindblom:1998wf} via
\begin{equation}
\label{alpha}
\alpha = 0.028 \left( \frac{h_0} {10^{-24}} \right) \left( \frac{\mbox{100 Hz}}
{f} \right)^3 \left( \frac{D} {\mbox{1 kpc}} \right),
\end{equation}
for a typical neutron star, with about a factor 2--3 uncertainty depending on
the mass and equation of state---see Eq.~(24) of \citet{Owen:2010ng} and the
discussion preceding it for details.
We used this equation to convert both the energy-conservation limit and the
direct 95\% confidence obtained here.
The results are plotted in the bottom panel of Fig.~\ref{f:uls2} for the Vela
Jr.\ wide search.
Like the plots of upper limits on fiducial ellipticity, the $\alpha$ upper
limit plots are tilted, curved versions of the $h_0$ upper limit plots.
Thus we do not display them for the other searches, although we do summarize
all of the ranges in Table~\ref{t:limits}.
Similarly to the case of fiducial ellipticity, the quantity most directly
inferred from $h_0$ here is the ($m=2$ part of the) current quadrupole.
While $\alpha$ is a convenient dimensionless measure, the conversion
factor---like that for $\epsilon$---is uncertain by a factor of a few.

\section{Discussion}
\label{s:disc}

\begin{table}
\begin{center}
\caption{
\label{t:long}
Upper limit data
}
\begin{tabular}{lcc}
\hline\hline
Search & Frequency (Hz) & $h_0$ upper limit \\
\tableline
G1.9+0.3 & 141.5 & $7.38\times10^{-25}$ \\
G1.9+0.3 & 142.5 & $7.08\times10^{-25}$ \\
G1.9+0.3 & 143.5 & $7.09\times10^{-25}$ \\
G1.9+0.3 & 144.5 & $7.44\times10^{-25}$ \\
$\cdots$ & $\cdots$ & $\cdots$ \\
\hline \\
\end{tabular}
\end{center}
This table lists data for our observational upper limits on $h_0$ for all
searches, i.e.\ the black points in Fig.~\protect\ref{f:uls1} and the top
panel of Fig.~\protect\ref{f:uls2}.
Frequencies are central frequencies for the upper limit bands.
Only a portion of this table is shown here to demonstrate its form and
content.
A machine-readable version of the full table is available.
\end{table}

Our searches improved sensitivity and parameter space coverage over previous
searches, beat indirect limits on \ac{GW} emission from electromagnetic
observations, and entered the range of theoretical predictions for neutron
stars.

The best direct (observational) upper limits on $h_0$ and the indirect
(theoretical) upper limits on $h_0$ from energy conservation are shown in
Table~\ref{t:limits}.
The \ac{S5} search for Cas~A \citep{S5CasA} obtained a best upper limit on
$h_0$ of $7\times10^{-25}$.
Our best \ac{S6} limit on Cas~A was $6\times10^{-25}$, less of an improvement
than the improvement in noise would indicate because we reduced the
integration time.
This in turn was because we searched a broader parameter space, including more
than doubling the frequency band.
Several of the \ac{S6} searches described here obtained upper limits on $h_0$
as strong (low) as $4\times10^{-25}$, nearly a factor of two better than
\citet{S5CasA} in spite of aiming in general for broad parameter space
coverage.
Several searches beat their corresponding indirect limits on $h_0$ by a factor
of two, and the Vela Jr.\ wide search beat its indirect limit by about a
factor of 20.

It is also interesting to compare our upper limits on neutron star fiducial
ellipticities and $r$-mode amplitudes to the maximum values predicted
theoretically.
Although these predictions have many uncertainties, observational limits that
beat them are more interesting than those that do not (even if the latter have
beaten energy conservation limits).
However we must be careful about interpreting \ac{GW} emission limits as
constraints on the physics of the targets---first of all, any target neutron
star may just be spinning too slowly to emit in the frequency band searched.
The spin frequency is half the \ac{GW} frequency for mass quadrupole emission,
or roughly three quarters for current quadrupoles ($r$-modes)---see
\citet{Idrisy2015} for a more precise range of numbers for the latter.
Each emission mechanism has a different range of predicted maximum
quadrupoles.
Depending on what is known about mechanisms for driving the quadrupoles toward
maximum, we may set some constraints on individual neutron stars, although we
cannot constrain universal properties such as the equation of state.

For mass quadrupoles supported by elastic forces, the analog of terrestrial
mountains, we can say this:
Many of our upper limits, summarized in Table~\ref{t:limits}, get well into
the range for stars composed of normal nuclear matter rather than exotic
alternatives, which has not been the case for previous \ac{GW} searches.
The most up-to-date numbers for elastically supported maximum quadrupoles are
in~\cite{JohnsonMcDaniel:2012wg}:
They correspond to maximum fiducial ellipticities of order $10^{-5}$ for
normal neutron stars, $10^{-3}$ for quark-baryon hybrid stars, and $10^{-1}$
for quark stars.
For instance the Vela Jr.\ wide search beat a fiducial ellipticity of
$10^{-5}$ over almost all of its frequency band.
Since little is known about what processes could drive an elastic quadrupole
toward its maximum in a young neutron star, we cannot use this information to
constrain the composition or other properties of the neutron star or the
properties of the processes.

Maximum values for magnetically supported mass quadrupoles depend on details
of the field configuration such as the relative strengths of the poloidal and
toroidal components as well as the hydrostatic structure of the star.
Although the literature on the problem grows rapidly, the highest
ellipticities predicted remain, as in \citet{S5CasA}, on the order of $10^{-4}
(B/10^{15}~\mathrm{G})^2$---see \citet{Ciolfi2013} for a recent example and
summary.
Magnetic fields must deform the star, and hence there is a minimum deformation
for a given internal field (found by varying the configuration) that does not
greatly differ from the maximum \citep{Mastrano2011}.
Thus our upper limits on $h_0$ correspond to upper limits on an average
internal magnetic field if the object is emitting \ac{GW} at the right
frequency---for example, about $10^{14}$~G for the Vela Jr.\ wide search over
much of its frequency band.
While these internal field limits are high (magnetar strength), they do not
require the objects to be magnetars since the external dipole fields could be
much lower \citep{Mastrano2011}.

It is also interesting to compare to the largest $r$-mode amplitudes predicted
by theory.
This is also a complicated subject, depending on the history as well as the
composition of the star.
As at the time of \citet{S5CasA}, the most detailed calculation of nonlinear
hydrodynamical saturation of the $r$-mode remains that of
\citet{Bondarescu2009}, and the answer is an amplitude of order $10^{-3}$ in
terms of the quantity $\alpha$ used here.
Thus, as seen in Fig.~\ref{f:uls2}, the Vela Jr.\ wide search reached
interesting values over most of its frequency band.
And as seen in Table~\ref{f:uls2}, most of the searches reached interesting
values at least at the high end of their frequency bands.
Since the \ac{GW}-driven instability of the $r$-modes drives them towards
saturation \citep{Andersson1998}, probably even in realistic conditions for
young neutron stars \citep{Lindblom:1998wf}, these upper limits also have more
constraining power than for elastic deformations:
If an object is emitting in the frequency band searched but not detected, we
can say that either the saturation amplitude is smaller or the damping
mechanisms more effective than commonly thought; though due to the complicated
physics of $r$-mode evolution scenarios it is difficult to be more precise.

In the near future, the Advanced \ac{LIGO} and Virgo interferometers will come
on-line and take data with strain noise amplitude reduced from \ac{S6} values
by a significant factor, which by the end of the decade will reach an order of
magnitude.
Re-running the analysis pipeline used here on such data would result in better
sensitivity to $h_0$, $\epsilon$, and $\alpha$ by the same factor.
Improved analysis methods are likely to improve the sensitivity even more,
making it interesting (i.e.\ possible to detect a signal or at least to set
upper limits that beat indirect limits) for many more supernova remnants and
other targets.

The authors gratefully acknowledge the support of 
the United States National Science Foundation (NSF) for the construction and operation of the LIGO Laboratory,
the Science and Technology Facilities Council (STFC) of the United Kingdom, 
the Max-Planck-Society (MPS), and the State of Niedersachsen/Germany 
for support of the construction and operation of the GEO600 detector,
the Italian Istituto Nazionale di Fisica Nucleare (INFN) and 
the French Centre National de la Recherche Scientifique (CNRS)
for the construction and operation of the Virgo detector. 
The authors also gratefully acknowledge research support from these agencies as well as by 
the Australian Research Council,
the International Science Linkages program of the Commonwealth of Australia,
the Council of Scientific and Industrial Research of India, 
Department of Science and Technology, India,
Science \& Engineering Research Board (SERB), India,
Ministry of Human Resource Development, India,
%the Istituto Nazionale di Fisica Nucleare of Italy, 
the Spanish Ministerio de Econom\'ia y Competitividad,
the Conselleria d'Economia i Competitivitat and Conselleria d'Educació, Cultura i Universitats of the Govern de les Illes Balears,
the Foundation for Fundamental Research on Matter supported by the Netherlands Organisation for Scientific Research, 
the Polish Ministry of Science and Higher Education, 
the FOCUS Programme of Foundation for Polish Science,
the European Union,
the Royal Society, 
the Scottish Funding Council, 
the Scottish Universities Physics Alliance, 
the National Aeronautics and Space Administration, 
the Hungarian Scientific Research Fund (OTKA),
the Lyon Institute of Origins (LIO),
the National Research Foundation of Korea,
Industry Canada and the Province of Ontario through the Ministry of Economic Development and Innovation, 
the National Science and Engineering Research Council Canada,
the Brazilian Ministry of Science, Technology, and Innovation,
the Carnegie Trust, 
the Leverhulme Trust, 
the David and Lucile Packard Foundation, 
the Research Corporation, 
and the Alfred P. Sloan Foundation.
The authors gratefully acknowledge the support of the NSF, STFC, MPS, INFN, CNRS and the
State of Niedersachsen/Germany for provision of computational resources. 

This research has made use of the SIMBAD database, operated at CDS,
Strasbourg, France.
This paper has been designated LIGO document number LIGO-P1400182-v6.

\appendix

\begin{table*}
\begin{center}
\caption{
\label{t:hwinj}
Nominal hardware injection parameters
}
\begin{tabular}{cccccccc}
\hline
\hline
Pulsar No.\ & RA+dec (J2000) & Base frequency (Hz) & -$\dot{f}$ (Hz/s) & $h_0$
& $\iota$ (rad) & $\psi$ (rad) & $\phi_0$ (rad)
\\
\tableline
0 & 044612.5$-$561303 & \hspace{0.5em}265.576360874\hspace{0.5em} &
$4.15\times10^{-12}$ & $2.47\times10^{-25}$ & 0.652 & \hspace{0.7em}0.770 &
2.66
\\
1 & 022934.5$-$292709 & \hspace{0.5em}849.029489519\hspace{0.5em} &
$3.00\times10^{-10}$ & $1.06\times10^{-24}$ & 1.088 & \hspace{0.7em}0.356 &
1.28
\\
2 & 142101.5$+$032638 & \hspace{0.5em}575.163548428\hspace{0.5em} &
$1.37\times10^{-13}$ & $4.02\times10^{-24}$ & 2.761 & $-$0.222 & 4.03
\\
3 & 115329.4$-$332612 & \hspace{0.5em}108.857159397\hspace{0.5em} &
$1.46\times10^{-17}$ & $1.63\times10^{-23}$ & 1.652 & \hspace{0.7em}0.444 &
5.53
\\
4 & 183957.0$-$122800 & 1398.60769871\hspace{1em} &
$2.54\times10^{-8}$\hspace{0.4em} & $4.56\times10^{-23}$ & 1.290 & $-$0.648 &
4.83
\\
5 & 201030.4$-$835021 & \hspace{1em}52.8083243593 &
$4.03\times10^{-18}$ & $4.85\times10^{-24}$ & 1.089 & $-$0.364 & 2.23
\\
6 & 235500.2$-$652521 & \hspace{0.5em}147.511962499\hspace{0.5em} &
$6.73\times10^{-9}$\hspace{0.4em} & $6.92\times10^{-25}$ & 1.725 &
\hspace{0.7em}0.471 & 0.97
\\
7 & 145342.1$-$202702 & 1220.77870273\hspace{1em} &
$1.12\times10^{-9}$\hspace{0.4em} & $2.20\times10^{-24}$ & 0.712 &
\hspace{0.7em}0.512 & 5.25
\\
8 & 232533.5$-$332507 & \hspace{0.5em}192.756892543\hspace{0.5em} &
$8.65\times10^{-9}$\hspace{0.4em} & $1.59\times10^{-23}$ & 1.497 &
\hspace{0.7em}0.170 & 5.89
\\
9 & 131532.5$+$754123 & \hspace{0.5em}763.847316497\hspace{0.5em} &
$1.45\times10^{-17}$ & $8.13\times10^{-25}$ & 2.239 & $-$0.009 & 1.01
\\
10 & 144613.4$+$425238 & \hspace{1em}26.3588743499 & $8.50\times10^{-11}$ &
$2.37\times10^{-24}$ & 2.985 & \hspace{0.7em}0.615 & 0.12
\\
11 & 190023.4$-$581620 & \hspace{1em}31.4248595701 &
$5.07\times10^{-13}$ & $1.80\times10^{-23}$ & 1.906 & \hspace{0.7em}0.412 &
5.16
\\
12 & 220724.6$-$165822 & \hspace{1em}39.7247751375 &
$6.25\times10^{-9}$\hspace{0.4em} & $2.66\times10^{-25}$ & 1.527 & $-$0.068 &
2.79
\\
\hline
\end{tabular}
\end{center}
Base frequencies are solar system barycentered at Jul 07, 2009 21:00:00 UTC
(the start of \ac{S6}).
The first derivatives $\dot{f}$ were constant, i.e.\ the injections did not
include second derivatives.
The inclination angle $\iota$, polarization angle $\psi$, and signal phase
offset $\phi_0$ were not used in this work.
They, and the detailed waveforms, are explained in detail in
\protect\citet{Jaranowski1998}.
\end{table*}

\ac{S6} featured a suite of hardware-injected continuous-wave signals, similar
to previous science runs.
Their nominal parameters (i.e.\ not allowing for calibration errors), in the
notation of \citet{Jaranowski1998}, are listed in Table~\ref{t:hwinj}.
They are used by most searches, including those described here, for basic
sanity checks of the analysis pipeline.
For each of the first ten, called Pulsars 0--9, we searched a 1~Hz wide band
around the injected frequency for a $T_\mathrm{span}$ of 10~days, and for
Pulsar~0 we also did a 20~day search (see below).
We did not search for Pulsars 10--12 since they were out of the frequency band
of the SFTs we used.
For each pulsar we ran the analysis pipeline using $f/|3\dot{f}|$ as the age
so that the search would cover the injected spin-down parameter in roughly the
middle of the range.

With these searches we were able to detect all ten hardware injections above
the ``further look'' threshold (95\% confidence in Gaussian noise).
Since Pulsar~0 was just barely above threshold in the first search, we made a
first follow-up by doubling the integration time to 20 days to verify that
$2\mathcal{F}$ doubled, similar to what would have been done in the early
stages of following up a plausible non-injected candidate.
The loudest injections (Pulsar~3 and Pulsar~8) triggered the \fscan\ veto,
which had to be switched off to complete this exercise.
Although this might cause concerns about the safety of the veto, these
injections are unreasonably loud, with $2\mathcal{F} \approx 2\times10^{4}$.
Real signals that loud would have been detected in earlier \ac{LIGO} data
runs.
Also, very few frequency bands triggered an \fscan\ veto in both detectors,
and we checked that (other than the loud hardware injections) these bands
corresponded to known instrumental artifacts.
By contrast, Pulsar~4 had $2\mathcal{F} \approx 2\times10^4$ and was not
\fscan-vetoed, apparently because of its large $|\dot{f}| >
2.5\times10^{-8}$~Hz/s spreading the power over several \ac{SFT} bins.

The recovered parameters of the hardware injections were typically off by the
amount expected from template parameter discretization and the fact that the
injections did not include a second spin-down parameter while the search
templates did.
In a real potential detection scenario, candidates would have been followed up
in a more sophisticated way, such as a hierarchical search or the gridless
method of \citet{ShaltevPrix}.

\bibliography{PostCasA}

\end{document}